\documentclass[leqno]{amsart}
\usepackage{amsmath}
\usepackage{amsfonts}
\usepackage{latexsym,amssymb,amsthm,amscd,epsf}

\def\e{\eta}
\def\a{\alpha}
\def\w{\omega}

\def\codim{\operatorname{codim}}
\def\const{\operatorname{const}}
\def\Hess{\operatorname{Hess}}

\def\NP{\mathcal{NP}}
\def\NS{\mathcal{NS}}
\def\Subset{\prec}

\numberwithin{equation}{section}
\newtheorem{theorem}{Theorem}[section]
\newtheorem{proposition}[theorem]{Proposition}
\newtheorem{corollary}[theorem]{Corollary}
\newtheorem{definition}[theorem]{Definition}
\newtheorem{conjecture}[theorem]{Conjecture}

\newtheorem{lemma}[theorem]{Lemma}

\newtheorem{example}[theorem]{Example}

\newtheorem{notation}[theorem]{Notation}

\def\CC{\mathbb{C}}
\def\RR{\mathbb{R}}

\def\RR{\mathbb{R}}

\def\al{\alpha}
\def\dist{\text {dist}}

\def\ep{\epsilon}

\def\<{\langle}
\def\>{\rangle}

\newcommand{\beal}{\begin{eqnarray}\begin{array}{l} }
\newcommand{\bear}{\begin{eqnarray}\begin{array}{r} }
\newcommand{\beac}{\begin{eqnarray}\begin{array}{c} }
\newcommand{\bealn}{\begin{eqnarray*}\begin{array}{l} }
\newcommand{\bearn}{\begin{eqnarray*}\begin{array}{r} }
\newcommand{\beacn}{\begin{eqnarray*}\begin{array}{c} }
\newcommand{\eea}{\end{array}\end{eqnarray}}
\newcommand{\eean}{\end{array}\end{eqnarray*}}

\newcommand{\beq}{\begin{equation} }
\newcommand{\eeq}{\end{equation} }

\begin{document}

\title{Mystery of point charges}


\author{Andrei Gabrielov$^\dag$}


\address{Department of Mathematics, Purdue University, W.Lafayette, IN
47907-1395, USA} \email{agabriel@math.purdue.edu}


\author{Dmitry Novikov$^\ddag$}


\address{Department of Mathematics, Purdue University, W.Lafayette, IN
47907-1395, USA} \email{dmitry@math.purdue.edu}

%    Information for third author

\author{Boris Shapiro$^\dag{}^\dag$}


\address{Department of Mathematics, University of Stockholm, S-10691, Sweden}
\email{shapiro@math.su.se}

\dedicatory{To Vladimir Igorevich  Arnold who taught us to study
classics}
%\mitya
\thanks{\\
  $^\dag$ Supported by NSF grants DMS-0200861 and
DMS-0245628\\
      $^\ddag$ Supported by NSF grants DMS-0200861\\
    $^\dag{}^\dag$  Boris Shapiro wants to acknowledge the hospitality
of the Department of
      Mathematics, Purdue University
during his visit in the Spring 2003.}

%    General info

 % Morse theory
%14N10\textbf{??? } % Voronoi diagrams

\date{\today}


\begin{abstract} We discuss the problem of finding an upper bound for the
number of equilibrium points of a potential of several fixed point
charges in $\RR^n$. This question goes back to J.~C.~Maxwell \cite
{Maxwell} and M.~Morse \cite {MorseCairns}. Using fewnomial theory we
show that for a given number of charges there exists an upper bound
independent of the dimension, and show it to be at most 12 for three
charges. We conjecture an exact upper bound for a given configuration
of nonnegative charges in terms of its Voronoi diagram,  and prove it
asymptotically.
\end{abstract}

\maketitle

\makeatletter

\renewcommand{\@evenhead}{\tiny \thepage \hfill  A.~GABRIELOV,
D.~NOVIKOV, B.~SHAPIRO \hfill}
\renewcommand{\@oddhead}{\tiny \hfill  MYSTERY OF POINT CHARGES
      \hfill \thepage}
\makeatother

\tableofcontents

\section{Introduction}

Consider a configuration of $l=\mu+\nu$ fixed point charges in
$\RR^n,\;n \ge 3$ consisting of $\mu$ positive charges with the
values $\zeta_{1},\ldots,\zeta_{\mu}$, and $\nu$ negative charges
with the values $\zeta_{\mu+1},\ldots,\zeta_{l}$. They create an
electrostatic field whose potential equals
\begin{equation}\label{pot1}
V(\bar x)=\left(\frac {\zeta_{1}}{ r_{1}^{n-2}}+\ldots+\frac
{\zeta_{\mu}}{ r_{\mu}^{n-2}}\right)+\left(\frac {\zeta_{\mu+1}}{
r_{\mu+1}^{n-2}}+\ldots+\frac {\zeta_{l}}{r_{l}^{n-2}}\right),
\end{equation}
where $r_{i}$  is the distance between the $i$-th charge and the
point $\bar x=(x_{1},\ldots,x_{n})\in \RR^n$ which we assume
different from the locations of the charges. Below we consider the
problem of {\bf finding effective upper bounds} on the number of
critical points of $V(\bar x)$, i.e. the number of points of
equilibrium of the electrostatic force. In what follows we mostly
assume that considered configurations of charges have only
nondegenerate critical points. This  guarantees that the number of
critical points is finite. Such configurations of charges and
potentials will be called $\textbf{nondegenerate}$. Surprisingly
little is known about this whole topic and the references are very
scarce.

In the case of $\RR^3$ one of the few known results  obtained by
direct application of Morse theory to $V(\bar x)$ is as follows, see
\cite{MorseCairns}, Theorem 32.1 and \cite {Kiang}, Theorem 6.

\begin{theorem}[Morse-Kiang]\label{MK}
Assume that the total charge $\sum_{i=1}^l \zeta_{j}$ in (\ref{pot1})
is negative (resp. positive). Let $m_{1}$ be  the number of the
critical points of index $1$ of $V$, and $m_{2}$ be  the number of
the critical points of index $2$ of $V$. Then $m_{2}\ge \mu$ (resp.
$m_{2}\ge \mu -1)$ and $m_{1}\ge \nu-1$ (resp. $m_{1}\ge \nu$).
Additionally, $m_{1}-m_{2}=\nu-\mu-1$.
\end{theorem}

Note that the potential $V(\bar x)$ has no (local) maxima or
minima due to its harmonicity.

\noindent
{\it Remark.} The remaining (more difficult) case  $\sum_{i=1}^\mu
          \zeta_{i}+\sum_{j=\mu+1}^l \zeta_{j}=0$ is treated in \cite
          {Kiang}.
       %   \end{remark}

    \noindent
{\it Remark.}
The above theorem has a generalization to any $\RR^n,\;$
$n\ge 3$ with $m_{1}$ being the number of the critical points of
index $1$ and $m_{2}$ being the number of the critical points of
index $n-1$.

%\end{remark}

\begin{definition} Configurations of charges with all nondegenerate
critical points and $m_{1}+m_{2}=\mu+\nu+1$  are called \textbf{
minimal}, see \cite {MorseCairns}, p. 292.

         \end{definition}

    \noindent
{\it Remark.}   Minimal configurations occur if one, for example,
places all         charges of the same sign on a straight line. On
the other hand, it is easy to construct generic nonminimal
configurations of charges, see \cite{MorseCairns}.

    \noindent
{\it Remark.} The major difficulty of this problem is that the {\bf
lower bound} on the number of  critical points of $V_n$ given by
Morse theory is known to be not exact. Therefore, since we are
interested in  an effective {\bf upper bound}, the Morse theory
arguments do not provide an answer.
      %\end{remark}

The question about the maximum (if it exists) of the number of points
of equilibrium of a nondegenerate configuration of charges in $\RR^3$
was posed in \cite {MorseCairns}, p. 293. In fact, J.~C.~Maxwell in
\cite{Maxwell}, section 113  made an explicit claim answering exactly
this question.

\begin {conjecture}[\cite{Maxwell}, see also \S\ref{sec:append} below] \label{Mx}
The total number of points of equilibrium (all assumed
nondegenerate) of any configuration with $l$ charges in $\RR^3$
never exceeds $(l-1)^2$.
          \end{conjecture}

\begin{figure}[!htb]
\centerline{\hbox{\epsfysize=5cm\epsfbox{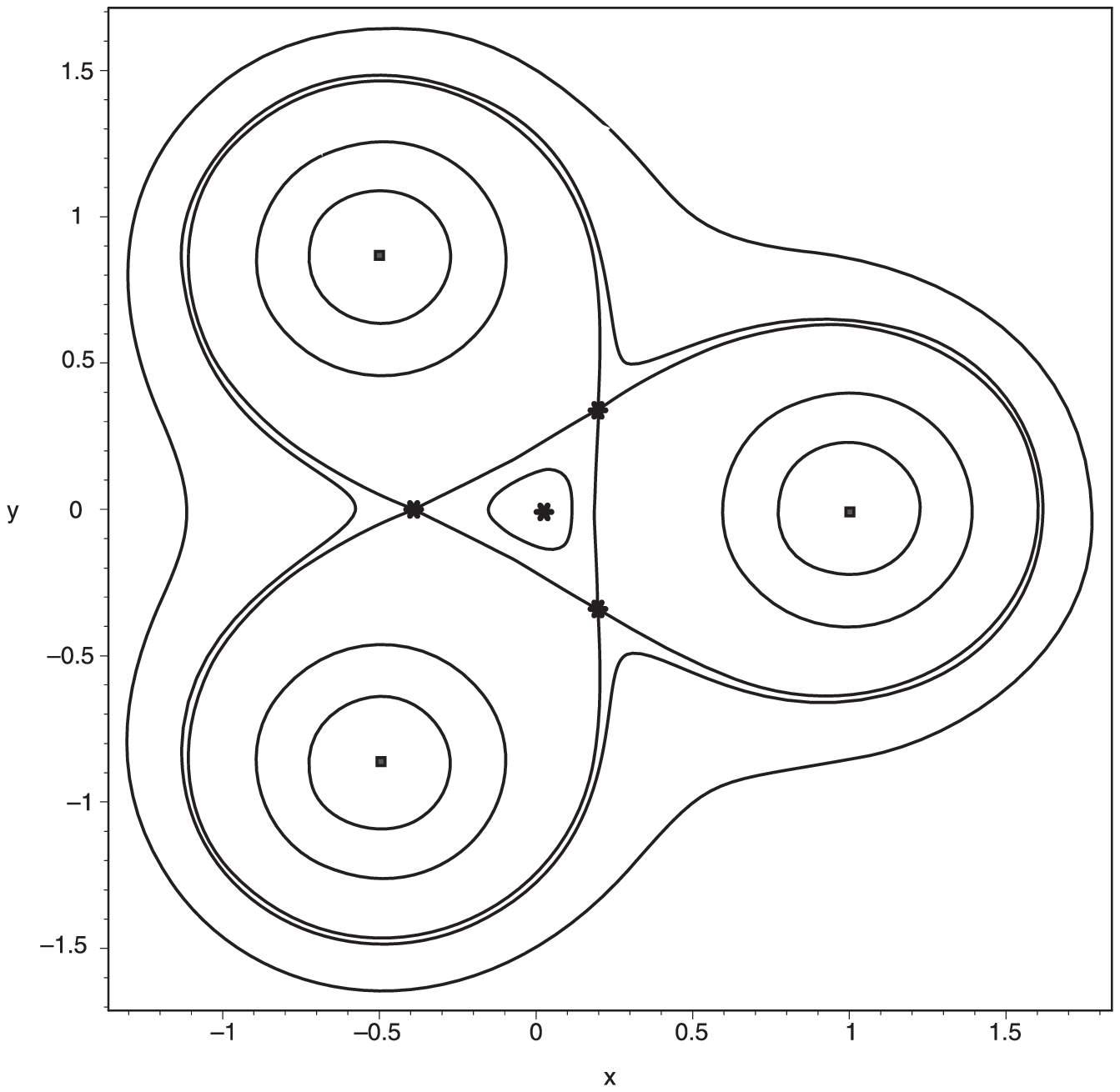}}
\hskip1cm\hbox{\epsfysize=5cm\epsfbox{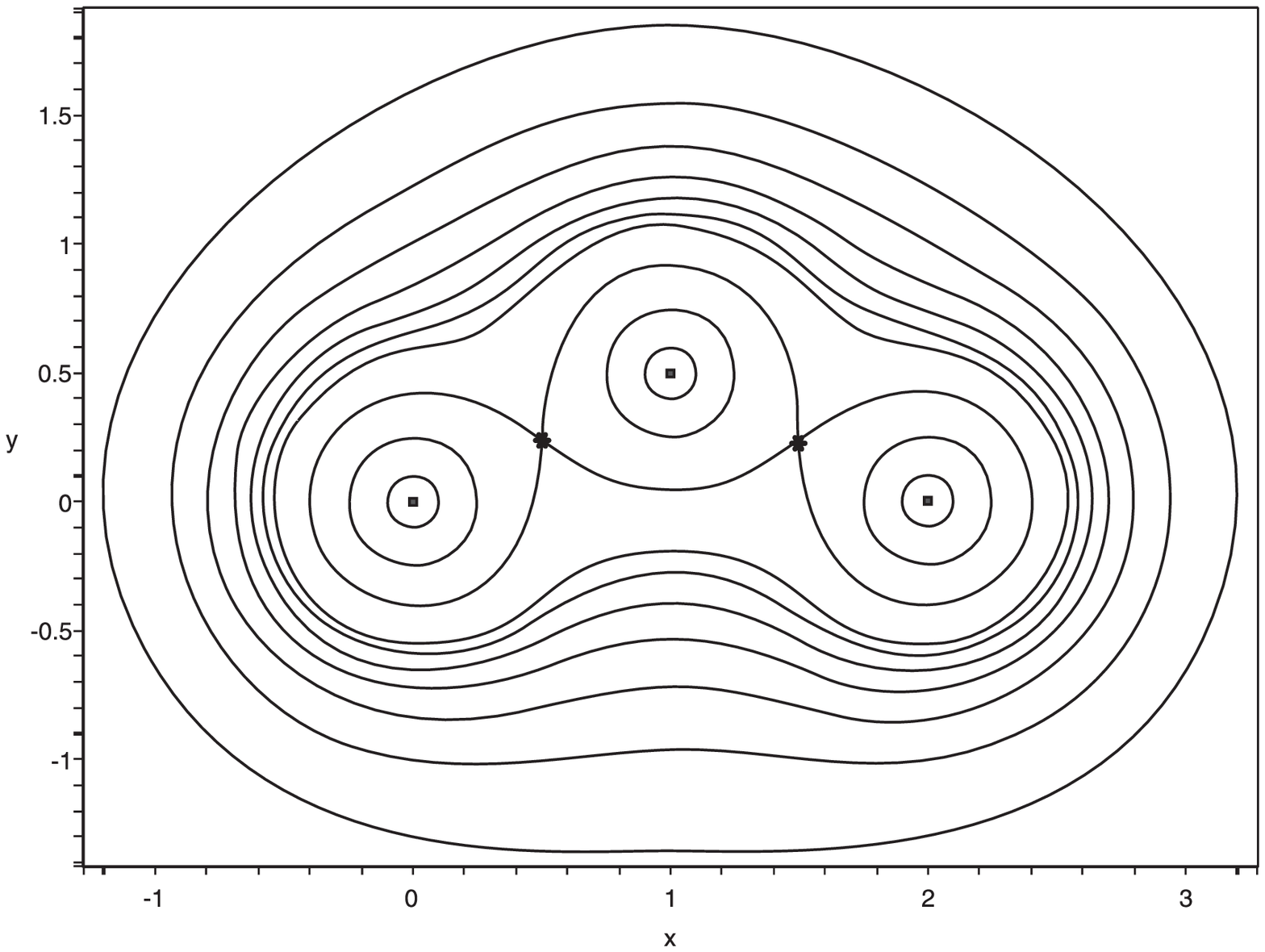}}}
\caption{Configurations with two and with four critical points.  }
\label{fig1}
\end{figure}

  \noindent
{\it Remark.} In particular,  there are at most $4$ points of
equilibrium for any configuration of $3$ point charges
according to Maxwell, see Figure~\ref{fig1}.
          %\end{remark}
%        \centerline{\bf But the above conjecture
%        remains unproved ever since.}

Before formulating our results and conjectures  let us first
generalize the set-up. In the notation of Theorem \ref{pot1}
consider the family of potentials depending on a parameter
$\alpha\ge0$ and given by

\begin{equation}\label{pot2}
V_{\alpha}(\bar x)=\left(\frac {\zeta_{1}}{
\rho_{1}^{\alpha}}+\ldots+\frac {\zeta_{\mu}}{
\rho_{\mu}^{\alpha}}\right)+\left(\frac {\zeta_{\mu+1}}{
\rho_{\mu+1}^{\alpha}}+\ldots+\frac
{\zeta_{l}}{\rho_{l}^{\alpha}}\right),
\end{equation}
where $\rho_{i}=r_{i}^2,\;i=1,\ldots,l$. (The choice of $\rho_{i}$'s
instead of $r_{i}$'s is motivated by convenience of algebraic
manipulations.)

\begin{notation} Denote by $N_{l}(n,\alpha)$ the maximal number
of the critical points  of the potential (\ref{pot2})  where the
maximum is taken over all nondegenerate configurations with $l$
variable point charges, i.e. over all possible values and
locations of
     $l$ point charges forming a nondegenerate configuration.
\end{notation}

Our first result is the following uniform  (i.e. independent on
$n$ and $\alpha$) upper bound.

\begin{theorem}\label{uni} a) For any $\alpha\ge0$ and any positive
integer $n$ one has
         \begin{equation}\label{unni}
        N_{l}(n,\alpha)\le 4^{l^2}(3l)^{2l}.
        \end{equation}

           b) For $l=3$ one has a significantly improved upper bound
      $$N_{3}(n,\alpha)\le 12.$$
         \end{theorem}

   \noindent
    {\it  Remark.}
         Note that the right-hand side of the formula (\ref{unni}) gives
         even for $l=3\;$ the horrible upper bound
         $139,314,069,504$.
      On the other hand, computer experiments suggest that Maxwell was right and that
      for any three charges there
         are at most 4 (and
         not $12$) critical points of the potential (\ref{pot2}), see
         Figure \ref{fig1}.
         %\end{remark}

\noindent {\it  Remark.} Figure \ref{fig1} shows the level curves of
the restrictions of the potential of three positive charges to the
plane they span in two essentially different cases (conjecturally,
the only ones).  The graph on the left has 3 saddles and 1 local
minimum and the graph on the right has just 2 saddle points.
%\end{remark}

\subsection{Voronoi diagrams and the main conjecture}\label{ssec:defs and main conj}

Theorem \ref{th:asymp} below determines the number of critical
points of the function $V_{\alpha}$ for large $\alpha$ in terms of
the combinatorial properties of the configuration of the charges. To
describe it we need to introduce several notions.

\medskip
\noindent\textbf{{Notation.}} By a (classical) \textbf{ Voronoi
diagram} \footnote {The first known application of Voronoi diagrams
can be traced back to Aristotle's De Caelo where Aristotle asked how
a dog faced with the choice of two equally tempting meals could
rationally choose between the two. These ideas were later developed
by the known French philosopher and physicist Jean Buridan
(1300-1356) who sowed the seeds of religious scepticism in Europe.
Buridan allowed that the will could delay the choice in order to more
fully assess the possible outcomes of the choice. Later writers
satirized this view in terms of an ass who, confronted by two equally
desirable and accessible bales of hay, must necessarily starve while
pondering a decision. Apparently the Roman Catholic Church found
unrecoverable errors in Buridan's arguments since about hundred and
twenty years after his death a posthumous campaign by Okhamists
succeeded in having Buridan's writings placed on the Index Librorum
Prohibitorum (List of Forbidden Books) from 1474-1481.} of a
configuration of pairwise distinct points (called sites) in the
Euclidean space $\RR^n$ we understand the partition  of $\RR^n$ into
convex cells according to the distance to the nearest site, see e.g.
\cite{Edel} and \cite{Prep}.

A \textbf{Voronoi cell} $S$ of the Voronoi diagram consists of all points
having exactly the same set of  nearest sites. The set of all
nearest sites of a given Voronoi cell $S$ is denoted by $\NS(S)$. One can
see that each Voronoi cell is a interior of a convex
polyhedron, probably of positive codimension. This is a
slight generalization of traditional terminology, which considers the
Voronoi cells of the highest dimension only.

A Voronoi cell of the  Voronoi diagram of a configuration  of sites is
called \textbf{effective} if it intersects the convex hull
of $\NS(S)$.

If we have an additional affine subspace $L\subset \RR^n$ we call
a Voronoi cell $S$ of the  Voronoi diagram of a configuration of charges in
$\RR^n$ \textbf{effective with respect to $L$} if $S$ intersects the
convex hull of the orthogonal projection of $\NS(S)$ onto $L$.

A configuration of points is called \textbf{ generic} if  any Voronoi
cell $S$ of its Voronoi diagram of any codimension $k$   has exactly
$k+1$ nearest cites and does not intersect the boundary of the convex
hull of $\NS(S)$.

%This implies that its closure does not intersect the convex hull of % $\NS(S)$ as well. TRUE, BUT NOT NEEDED.

A subspace $L$ intersects a Voronoi diagram \textbf{generically } if
it intersects all its Voronoi cells transversally, any Voronoi cell $S$  of
codimension $k$ intersecting $L$ has exactly $k+1$ nearest sites,
and $S$ does not intersect the boundary of the convex hull of
the orthogonal projection of $\NS(S)$ onto $L$.

The \textbf{ combinatorial complexity} (resp. \textbf{effective
combinatorial complexity}) of a given configuration of
points is the total number of cells (resp. effective cells) of all
dimensions in its Voronoi diagram.

\medskip
\noindent{\it Example.} Voronoi diagram of three non-collinear
points $A,B,C$ on the plane consists of seven Voronoi cells:
\begin{enumerate}
\item three two-dimensional cells $S_A,S_B,S_C$ with $\NS(S)$ consisting of one
point,
\item three one-dimensional cells $S_{AB},S_{AC},S_{BC}$ with $\NS(S)$ consisting of two
points.  For example, $S_{AB}$ is a part of a perpendicular bisector of the
segment $[A,B]$.
\item one zero-dimensional cell $S_{ABC}$ with $\NS(S)$ consisting of all three
points. This is a point equidistant from all three points.
\end{enumerate}

There are two types of generic configurations, see Figure \ref{fig3}.
First type is of an acute triangle
$\Delta ABC$ and then all Voronoi cells  are effective. Second type is of an obtuse
triangle $\Delta ABC$, and then (for the obtuse angle $A$) the  Voronoi cells $S_{BC}$ and
$S_{ABC}$ are not effective.

The case of the equilateral triangle $\Delta ABC$ is non-generic: the cell
$S_{ABC}$, though effective, lies on the boundary of the triangle.

\medskip
The following result motivates our main conjecture \ref{conj:main}
below.

\begin{theorem}\label{th:asymp}
%\text{}\\
\begin{itemize}
\item[a)]For any generic configuration of point charges of the same
sign there exists $\alpha_{0}>0$ such that for any $\alpha\ge
\alpha_{0}$ the  critical points of the potential $V_{\alpha}(\bar
x)$ are in one-to-one correspondence  with effective cells of
positive codimension in the Voronoi diagram of the considered
configuration. The Morse index of each critical point coincides with
the dimension of the corresponding Voronoi cell.

\item[b)]Suppose that an affine subspace $L$ intersects generically
the Voronoi diagram of a given configuration of point charges of
the same sign.

Then there exists $\alpha_{0}>0$ (depending on the configuration and
$L$) such that for any $\alpha\ge \alpha_{0}$ the critical points of
the restriction of the potential $V_{\alpha}(\bar x)$ to $L$ are in
one-to-one correspondence  with effective w.r.t. $L$ cells of
positive codimension in the Voronoi diagram of the considered
configuration. The Morse index of each critical point coincides with
the dimension of the intersection of the corresponding Voronoi cell with
$L$.
\end{itemize}
\end{theorem}

\medskip
Finally, our computer experiments in one- and two-dimensional
cases led us to the following optimistic

\begin{conjecture}\label{conj:main}
     \begin{itemize}
\item[a)]For any generic configuration of point charges of the same
sign and any $\alpha\ge \frac 1 2$ one has
\begin{equation}
a_{\al}^j\le \sharp^j,
\label{Morse}
\end{equation}
where  $a_{\al}^j$ is the number of the critical points of index $j$
of the potential $V_{\alpha}(\bar
x)$  and  $\sharp^j$ is the number of all effective Voronoi cells of
  dimension $j$ in the Voronoi diagram of the considered
configuration.

\item[b)]For any affine subspace $L$ generically
intersecting
the Voronoi diagram of a given configuration of point charges of
the same sign one has
\begin{equation}
a_{\al,L}^j\le \sharp^j_{L},
\label{RelMorse}
\end{equation}
where  $a_{\a,L}^j$ is the number of the critical points of index $j$
of the potential $V_{\alpha}(\bar x)$  restricted to $L$ and
$\sharp^j_{L}$ is the number of all Voronoi cells with $\dim (S\cap
L)=j$ effective w.r.t $L$ in the Voronoi diagram of the considered
configuration.

\end{itemize}

\end{conjecture}

We will refer to the inequality (\ref{Morse}) resp. (\ref{RelMorse})
as {\bf Maxwell} resp. {\bf relative Maxwell} inequality.

\medskip
\noindent
{\it Remark. } Theorem \ref{th:asymp} and Conjecture \ref{conj:main} were
inspired by two observations.  On one hand, one can compute the
limit of a properly normalized potential $V_{\al}(\bar x)$ when
$\alpha\to\infty$. Namely, one can easily show that
$$
\lim_{\alpha\to\infty}{V_{\alpha}}^{-\frac 1 \al}(\bar
x)=V_{\infty}(\bar x)=\min_{i=1,\ldots,l} {\rho_i}(\bar x).
$$
This limiting function is only piecewise smooth. However, one can
still define critical points of $V_{\infty}(\bar x)$  and their Morse
indices. Moreover, it turns out that for generic configurations every
critical point of $V_{\infty}(\bar x)$ lies on a separate effective
cell of the Voronoi diagram whose dimension equals the Morse index of
that critical point, see \ref{cr:infty}.   Theorem \ref{th:asymp}
above claims that for sufficiently large $\alpha$ the situation is
the same, except that the critical point does not lie exactly on the
corresponding Voronoi cell (in fact, it lies on $O(\alpha^{-1})$
distance from this Voronoi cell, see Lemma \ref{lem1:dim=0} and
\ref{lem1:dim=k}). On the other hand, computer experiments show that
the largest number of critical points (if one fixes the positions and
values of charges) occurs when $\alpha\to\infty$.

Even the special case of the conjecture \ref{conj:main} when $L$ is
one-dimensional is of interest and still open. Its slightly stronger
version supported by extensive numerical evidence can be
reformulated  as follows.
\begin{conjecture}\label{conj:1dim}
Consider an $l$-tuple of points
$(x_{1},y_{1}),\ldots,(x_{l},y_{l})$ in $\RR^2$. Then for any
values of charges $(\zeta_{1},\ldots,\zeta_{l})$ the function
$V_{\al}^*(x)$ in (one real) variable $x$ given by
\begin{equation}
V_{\al}^*(x)=\sum_{i=1}^l\frac{\zeta_{i}}{((x-x_{i})^2+y_{i}^2)^\alpha}
\end{equation}
has at most $(2l-1)$ real critical points, assuming
$\alpha \ge \frac 1 2$.
\end{conjecture}

\noindent
{\it Remark.}
In the simplest possible case $\alpha=1$ conjecture
\ref{conj:1dim} is equivalent to showing that real polynomials of
degree $(4l-3)$ of a certain form have  at most $(2l-1)$ real
zeros.
%\end{remark}

\subsubsection{Complexity of Voronoi diagram and Maxwell's conjecture}

In the classical planar case one can show that the total number of
cells of positive codimension of the Voronoi diagram of any $l$
sites on the plane is at most $5l-11$ and this bound is exact.

Since $(l-1)^2$ is larger than the conjectural exact upper bound
$5l-11$ for all $l>5$ and coincides with $5l-11$ for $l=3,4$, we
conclude that Conjecture~\ref{conj:main} implies a stronger form of
Maxwell's conjecture for any $l$ positive charges on the plane and
any $\al\ge \frac 1 2$.

For $n>2$ the worst-case complexity $\Gamma(l,n)$ of the classical
Voronoi diagram of an $l$-tuple of points in $\RR^n$ is
$\Theta(l^{[n/2+1]})$, see \cite {Edel}. Namely, there exist
positive constants $A<B$ such that $A l^{[ n/2+1]}<\Gamma(l,n)< B
l^{[n/2+1]}$. Moreover, the Upper Bound Conjecture of the convex
polytopes theory proved by McMullen implies that the number of
Voronoi cells of dimension $k$ of a Voronoi diagram of $l$ charges in
$\RR^n$ does not exceed the number of $(n-k)$-dimensional faces in
the $(n+1)$-dimensional cyclic polytope with $l$ vertices, see
\cite{GoodmanRourke, McMullen}. This bound is exact, i.e. is
achieved for some configurations, see \cite{Seidel}.

In $\RR^3$ this means that the number of $0$-dimensional Voronoi cells of
the Voronoi diagram of $l$ points is at most $\frac{l(l-3)}{2}$, the
number of $1$-dimensional Voronoi cells is at most $l(l-3)$,  and the
number of $2$-dimensional Voronoi cells is at most $\frac{l(l-1)}{2}$.

We were unable to find a similar result about the number of {\em
effective} cells of Voronoi diagram. However, already for a regular 
tetrahedron the number of effective cells is 11, which is greater
than the Maxwell's bound 9. Thus a stronger version of Maxwell's
conjecture in $\RR^3$ fails: the number of critical points of
$V_{\alpha}$ could be bigger than $(l-1)^2$ for $\alpha$ sufficiently
large.

However Maxwell's original conjecture miraculously agrees with the
Maxwell inequalities (\ref{Morse}) and  we obtain the following
conditional statement.

\begin{theorem}
Conjecture~\ref{conj:main} implies the validity of the original
Maxwell's conjecture  for any configuration of positive charges in
$\RR^3$ in the standard $3$-dimensional Newton potential, i.e.
$\al=\frac 1 2$. \label{th:origMx}
     \end{theorem}

\subsubsection*{Existing literature and
acknowledgements.} Logarithmic potentials in $\RR^3$ similar to
(\ref{pot2})  (i.e. the case of the electrostatic force proportional
to the inverse of the distance) were studied in a number of papers of
J.~L.~Walsh, see \cite{MotzkinWalsh} and references therein. In this
case it is possible to generalize the classical Gauss-Lucas theorem
and some results on Jensen's circles for polynomials in one complex
variable to real vector spaces of higher dimension.

Critical points of a logarithmic potentials in $R^2=C$ are zeros of
some univariate polynomial of degree $\le l-1$, so the upper bound for the
number of critical points is $l-1$, see \cite{Marden}.

Some interesting examples of electrostatic potentials whose critical
points form curves were considered in \cite{Janu}. The question
whether degenerate electrostatic potential defined by a finite number
of charges can have an analytic arc of critical points was stated in
\cite{MorseCairns}, p.294. Finally, the results about instability of
critical points for more general potentials and dynamical systems are
obtained in \cite {Kozlov}. Instability in our context follows from
subharmonicity of the considered potential and was already mentioned
in \cite{Maxwell}, section 116 under the name Earnshaw's theorem.

The structure of the paper is as follows. In \S \ref{sec:proofs}
we prove the above results and present the computer evidence for
our main conjecture. \S\ref{sec:rems} contains further remarks and
open problems related to the topic. Finally, in \S\ref{sec:append}
we reproduce the original section 113 of \cite{Maxwell} where
Maxwell presents the arguments of Morse theory (developed at least
50 years later), and names  the ranks of the 1st and the 2nd
homology groups of domains in $\RR^3$ in the language of
(apparently existing) topology of 1870's to formulate his claim.

The authors are sincerely grateful to A.~Eremenko, A.~Fryntov,
D.~Khavinson, H.~Shapiro, M.~Shapiro and A.~Vainshtein for valuable discussions
and references.

       \label{sec:intro}

        \section{Proofs}
     \label{sec:proofs}

We start this section with a discussion of the nondegeneracy
requirement and the (co)dimension of the affine span of a
configuration of point charges.

\subsection{Relation between number of charges and dimension}

Consider a nondegenerate configuration of $l=\mu+\nu$ point charges
in $\RR^n$, and let $L\subseteq \RR^n$ be the affine subspace spanned
by the points where the charges are located. Evidently, $\dim L\le
l-1$.

\begin{theorem}\label{pr:l-1}
If all critical points of the potential $V_{\alpha}$ are isolated,
then either all critical points belong to $L$ or $n\le l-1$.
\end{theorem}

Let us first  show that it is enough to consider
the cases $n\le l$ only.
\begin{lemma}
If a configuration of charges in $\RR^n$ has only isolated critical
points then either all its critical points belong to $L$  or $L$ is a
(real) hyperplane in $\RR^n$. \label{lm:dim}
\end{lemma}

\begin{proof}
Indeed, assume that there is a critical point outside $L$ and $\codim L>1$. Then the whole orbit of
this point under the action of the group of rotations of
$\mathbb{R}^n$ preserving $L$ consists of critical points
(since this action preserves the potential).
\end{proof}

To complete the proof one has to exclude the case $n=l$. We show that
if $\dim L=l-1$ then all critical points of the potential are in $L$.

\begin{lemma}\label{lm:simplexprep}
If one can find a hyperplane $H$ in $L$
separating positive charges from the negative ones,  then the
potential of the configuration has no critical points outside
$L$.
\end{lemma}

\noindent
\begin{proof} Indeed, let $x\not \in L$ be any point outside $L$,
and let $H_x$ be any hyperplane containing both $x$ and $H$ and
transversal to $L$. Let $n$ be a vector normal to $H_x$ at $x$. The
the signs of scalar products of the gradients of the potentials of
each charge with $n$ are the same, so $x$ cannot be an equilibrium
point.
\end{proof}

\begin{corollary}
Potential of any configuration of positive charges has no critical
points outside $L$.
\end{corollary}

\begin{corollary}\label{lm:simplex}
Any configuration with $l$ point charges such that $\dim L=l-1$ has
no critical points of potential outside $L$.
\end{corollary}
\begin{proof}
These points should form a non-degenerate simplex, and any subset of
vertices of a simplex can be separated from the rest of the vertices
by a hyperplane, so the claim follow from the previous Lemma.
\end{proof}

\noindent{\em Remark.\/} As one can see from the proof, the set of
critical points of a configuration is a union of spheres with centers
in $L$ and of dimension equal to $\codim L$. As $n$ grows, the change
of the dimension of spheres is the only parameter that changes, so
the case $\codim L=1$ is the most general one.

We conclude that in any case it is enough to consider the case $n\le
l-1$.
\medskip

\subsection {Proof of Theorem \ref{uni}.a}

The proof is an application of the theory of fewnomials developed
by A.~G.~Khovanskii in \cite{Khovanskii}. A serious drawback of
this theory is that the obtained estimates, though effective, are
usually highly excessive. Applying the methods, rather than the
results of this theory one might get a much better estimate which
we illustrate while proving Part b) of Theorem \ref{uni}.

We start with the following result from   \cite[\S1.2]{Khovanskii}.

\begin{theorem}[Khovanskii]\label{exppoly}
Consider a system of $m$ \emph{quasipolynomial} equations
$$
P_1(\bar{u},\bar{w}(\bar{u}))=...=P_m(u,\bar{w}(\bar{u}))=0,\quad \bar{u}=(u_1,...,u_m),
$$
where each $P_i$ is a real polynomial of degree $d_i$ in $(m+k)$
variables $(u_1,...,u_m,w_1,...,w_k)$ and
$$w_j=\exp\langle \bar{a}_j,\bar{u}\rangle,\qquad \bar{a}_j=(a_j^1,...,a_j^m)\in\RR^m,\quad j=1,...,k.$$

Then the  number of real isolated solutions of this system does
not exceed
$$
d_1\cdots d_m\left(d_1+\dots+d_m+1\right)^k2^{k(k-1)/2}.
$$
\end{theorem}

The estimate of Theorem \ref{uni}.a will follow from a
presentation of the critical points of a configuration of point
charges as solutions of an appropriate system of quasipolynomial
equations, see below.

\subsubsection{Constructing a quasipolynomial system.}

Consider a configuration with $l$ point charges in $\RR^n$. Denote
by $\bar x=(x_1,...,x_{n})$ the coordinates of a critical point and
denote by $(c_1^i,...,c_{n}^i),\;i=1,\ldots,l$ the coordinates of
the $i$-th charge. We assume that the $1$-st charge is placed at the
origin, i.e. that $c_1^1=\dots=c_n^1=0$.

The first $l$ equations of our system define  the indeterminates
$\bar\rho=(\rho_1,\ldots,\rho_{l})$ as  the squares of distances
between the variable point $\bar x$ and the charges. They can be rewritten as
\begin{equation}\label{sys1}
P_1(\bar x,\bar \rho)=...=P_{l}(\bar x,\bar\rho)=0,
\end{equation}
where
\begin{equation}
P_1(\bar x,\bar \rho)=\sum_{j=1}^{n} x_j^2-\rho_1, \quad  %\label{sys1A}\\
P_i(\bar x,\bar \rho)=\rho_1 -\rho_i
+\sum_{j=1}^{n}c_j^i(2x_j+c_j^i),\quad i=2,...,l. \label{sys1B}
\end{equation}

The second group of equations expresses the fact that the point
$\bar x=(x_1,...,x_{n})$ is the critical point of the potential
$V_{\al}(\bar x)=\sum \zeta_i \rho_i^{-\alpha}$. Namely,
$$
\frac{\partial}{\partial x_j}V_{\al}(\bar x)=\sum_{i=1}^l \zeta_i
\frac{\partial}{\partial x_j}\rho_i^{-\alpha}=-2\alpha\sum_{i=1}^l
\zeta_i v_i(x_j-c_j^i)=P_{l+j}(\bar x,\bar v),\quad j=1,...,n
$$
where we denote
\begin{equation}\label{sysadd2}
v_i=\rho_i^{-\alpha-1},\quad i=1,...,l\quad\text{and \quad} \bar
v=(v_{1},\ldots,v_{l}).
\end{equation}

Introducing variables $s_i=\log \rho_i$ we get the  system:
$$
P_1(\bar x,\bar s,\bar \rho,\bar v)=...=P_{n+l}(\bar x,\bar s,\bar
\rho,\bar v)=0,
$$
of $(l+n)$ quasipolynomial equations in $(l+n)$ variables $(\bar x,\bar s,\bar
\rho,\bar v)$ with
$$
\rho_i=\exp(s_i),\qquad v_i=\exp(-(\alpha+1) s_i/2),\qquad
i=1,...,l.
$$

This system has the type described in Theorem \ref{exppoly}, with
$m=n+l,$\; $k=2l,$\; $\deg P_1=\deg P_{l+1}=...=\deg P_{n+k}=2,$
and $\deg P_2=...=\deg P_l=1$. By Proposition
\ref{pr:l-1} one has  $n\le l-1$ which implies the required
estimate:
$$
N_{l}(n,\alpha)\le N_{l}(l-1,\alpha)\le
4^{l^2}9^{l}l^{2l}=4^{l^2}(3l)^{2l}.
$$
For example, for $l=3$ one gets $N_3(n,\alpha)\le
139,314,069,504.$

\subsection{Proof of  Theorem \ref{uni}.b}

As we mentioned above, one can do much better by applying the
fewnomials method rather than results, and here we demonstrate this
in the case of three charges. By Proposition \ref{pr:l-1} we can
restrict our consideration to the case of $\RR^{l-1}=\RR^2$. We also
use coordinates $(x,y)$ instead of $(x_1,x_2)$.

The scheme of this rather long proof is as follows. We make a change
of variables, passing from $(x,y)$ to new variables  $(f,g)$. In the
new coordinates the equilibrium points coincide with the intersection
points  of two explicitly written planar curves $\gamma_1$ and
$\gamma_2$ in the positive quadrant $\RR_+^2$ of the real plane. Both
curves are separating solutions of Pfaffian forms. The
Rolle-Khovanskii theorem applied twice produces two real polynomials
$R$ and $Q$ such that the required upper bound can be given in terms
of the number of their common zeros in $\RR_+^2$. The latter is
bounded from above by the Bernstein-Kushnirenko bound minus the
number of common roots of $R$ and $Q$ lying outside $\RR^2_+$.

\subsubsection{Changing variables and getting  system of equations}
To emphasize that the methods given below can be generalized we make
the change of variables in the situation of $l$ charges in $\RR^{n}$
(we assume, as before, that $n\le l-1$). This, as a byproduct,
produces another proof of Theorem \ref{uni} with a somewhat better
upper bound.

As above, we assume that the charges $\zeta_i$ are located at
$(c^i_1,...c^i_n)$, $i=1,...,l$, and consider the potential
$$
V_{\alpha}(x_1,...,x_{n})=\sum_{i=1}^l\zeta_i\rho_i^{-\alpha},
\qquad\text{where}\quad\rho_i=\sum_{j=1}^{n}(x_j-c^i_j)^2,\quad
i=1,\dots,l.
$$

The system of equations defining the critical points of
$V_{\alpha}(\bar x)$ is
$$
\frac{\partial V_{\alpha}(\bar x)}{\partial x_j}=0,\quad j=1,...,n,
\qquad\text{where}\quad\frac{\partial V_{\alpha}(\bar x)}{\partial
x_j}=-2\alpha\sum_{i=1}^l\zeta_i\rho_i^{-\alpha-1}(x_j-c^i_j).
$$

Introducing $h_i=\rho_i^{-\alpha-1}$  one can solve each
equation of this system and express $x_j$ in terms of $h_i$:
\begin{equation}\label{xs in hs}
x_j=\frac{\sigma_j}{\sigma},\qquad\text{where}\quad
\sigma=\sum_{i=1}^l\zeta_i h_i,\quad \sigma_j=\sum_{i=1}^l \zeta_i
h_i c^i_j
\end{equation}
are homogeneous linear functions of $h_i$. The equilibrium points
correspond to the solutions of the following system of equations
obtained from the definition of $h_i$ by substitution of
$\sigma_j/\sigma$ instead of $x_j$:
\begin{equation}\label{system:general}
h_i^{-\frac{1}{\alpha+1}}=\frac{\xi_i}{\sigma^2},\qquad\text{where}\quad
\xi_i=\sum_{j=1}^{n}(\sigma_j-c_j^i\sigma)^2, \quad i=1,...,l.
\end{equation}

This system has following remarkable properties:
\begin{proposition}\label{common zero}
a) Any solution of $\sigma=\xi_1=0$ is also a zero of all
$\xi_i$'s;\\
b) each $\xi_i$ is a strictly positive real quadratic polynomial
independent of $h_i$.
\end{proposition}
\begin{proof}Indeed,
$\xi_i-\xi_1=\sigma\sum_{j=1}^{n}(c_j^1-c_j^i)(2\sigma_j-\sigma(c_j^i+c_j^1))$.

The second statement is evident except the independence on $h_i$,
which is proved by direct computation.\end{proof}

\noindent {\it Remark.} Note that the above system can be
represented as a system of quasipolynomials as in Theorem
\ref{exppoly}. Namely, the equations in (\ref{system:general}) are
polynomials in $h_i, h_i^{1/(\alpha+1)}$. Introducing  $s_i=\log
f_i$ one can apply Theorem \ref{exppoly}. After several small tricks
-- dehomogenization of the system, introduction of a new variable
$z=\xi_1$ and noting that the expression for $\xi_1-\xi_i$ becomes then linear -- we obtain
an upper bound $2\cdot4^{l^2}(2l+3)^{2l}$ on the number of
equilibrium points of a system of $l$ charges. For $l>3$ this bound
is somewhat better  than the bound \ref{unni}.
%\end{remark}

\medskip
Now, let us use the previous construction for $l=3$ and $n=2$.
Without loss of generality we can assume that the three charges with
the values $\zeta_1, \zeta_2, 1$ are located at $(0,0)$, $(1,0)$ and
$(a,b)$ respectively.

Expressions  (\ref{xs in hs}) are homogeneous in $h_j$, so we
introduce the nonhomogeneous variables $f$ and $g$ as follows
\begin{equation}
f=\frac{h_2}{h_1}=\left(\frac{\rho_1}{\rho_2}\right)^{\a+1}\quad\text{and}\quad
g=\frac{h_3}{h_1}=\left(\frac{\rho_1}{\rho_3}\right)^{\a+1}.
\label{E:fgdefs}
\end{equation}
Then equations (\ref{xs in hs}) become:
\begin{equation}\label{xyinfg}
x=\frac{ag+\zeta_2f}{\zeta_1+\zeta_2f+g},\qquad
y=\frac{bg}{\zeta_1+\zeta_2f+g},
\end{equation}

The system (\ref{system:general}) reduces to the following two
equations describing
two curves $\gamma_1$ and $\gamma_{2}$ in the positive quadrant $\RR^2_+$ of the $(f,g)$-plane:
\begin{eqnarray}\label{gammaidefs}
\quad\gamma_1=\left\{f^{1/(\alpha+1)}\xi_2\xi_1^{-1}=1\right\}, \quad
\gamma_2=\left\{g^{-1/(\alpha+1)}\xi_2=f^{-1/(\alpha+1)}\xi_3\right\}.
\end{eqnarray}
Here
\begin{eqnarray*}
\xi_1&=&(ag+\zeta_2f)^2+b^2g^2,\\
\xi_2&=&((a-1)g-\zeta_1)^2+b^2g^2,\\
\xi_3&=&((a-1)\zeta_2f+a\zeta_1)^2+b^2(\zeta_2f+\zeta_1)^2.
\end{eqnarray*}

  The following  facts about $\xi_i$ follow from the Proposition \ref{common zero}:
\begin{proposition}\label{prop:common zero in 3dim}
\begin{enumerate}
\item $\xi_2$ depends only on $g$, and $\xi_3$ depends only on $f$;
\item $\xi_2,\xi_3$ are strictly positive quadratic polynomials;
\item $\xi_1$ is a positive definite homogeneous quadratic form;
\item $\xi_1=\xi_2=\xi_3=0$ have two complex solutions.
\end{enumerate}
\end{proposition}

The goal of all subsequent computations is to give an upper bound on
the number $N$ of the points of intersection of $\gamma_1$ and
$\gamma_2$ in $\RR_+^2$. We are able to obtain the following
estimate proved below.

\begin{proposition} The number of intersection  points of $\gamma_1$ and
$\gamma_2$ lying in $\RR_+^2$ is at most 12. \label{pr:12}
\end{proposition}

Note, that any intersection point of $\gamma_1$ and $\gamma_2$ lying
in the positive quadrant $\RR_+^2=\{f>0,g>0\}$ corresponds, via
(\ref{xyinfg}), to a unique critical point of $V_{\al}(x,y)$, so the
estimate of \ref{uni}.b immediately follows.

\subsubsection{Rolle-Khovanskii theorem} Before we move further let us
recall the $\RR^2$-version of a generalization of Rolle's theorem
due to Khovanskii.

Suppose that  we are given a smooth differential 1-form $\w$
defined in a domain  $D\subset {\mathbb R}^2$. Let $\gamma\subset D$ be a
(not necessarily connected) one-dimensional integral submanifold
of  $\w$.

\begin{definition}
We say that $\gamma$ is a {\bf separating solution of $\w$}
  if
\begin{itemize}
\item[a)] $\gamma$ is the boundary of some (not necessarily
connected) domain $U$;
\item[b)] the coorientations of $\gamma$ defined by $\w$ and by $U$ coincide
(i.e. $\w$ is positive on the outer normal to the boundary of
$U$).
\end{itemize}
\end{definition}

Let $\gamma_1$, $\gamma_2$ be two separating solutions of two
$1$-forms $\w_1$ and $\w_2$ resp.

\begin{theorem}[see \cite{Khovanskii}]

     $$  \sharp(\gamma_1,\gamma_2)\le \sharp(\gamma_1)+
     \flat(\gamma_1,\gamma_2),$$
     where $\sharp(\gamma_1,\gamma_2)$ is the number of intersection points of
$\gamma_1$ and $\gamma_2$, $\sharp(\gamma_1)$ is the number of
non-compact components of $\gamma_1$ and
$\flat(\gamma_1,\gamma_2)$ is the number of the points of contact
of $\gamma_1$ and $\w_2$, i.e. the number of points of $\gamma_1$
such that $\w_2(\dot{\gamma_1})=0$. (One can also characterize the
latter points as the intersection points of $\gamma_1$ with an
algebraic set $\{\w_1\wedge\w_2=0\}$.) \label{th:Khov}
\end{theorem}

\subsubsection{First application of Rolle-Khovanskii theorem}

We apply Theorem \ref{th:Khov} to the curves $\gamma_1,\gamma_2$
defined in (\ref{gammaidefs}). These curves are integral curves in
$\RR_+^2$ of the one-forms $\eta_1$ and $\eta_2$ respectively, where
\begin{eqnarray}
\eta_1&=&\frac{df}{(\alpha+1)f}+\frac{\xi_2'dg}{\xi_2}-\frac{d\xi_1}{\xi_1},
\\
\eta_2&=&\left(-\frac{1}{(\alpha+1)f}+\frac{\xi_3'}{\xi_3}\right)df+\left(\frac{1}
{(\alpha+1)g}-\frac{\xi_2'}{\xi_2}\right)dg.
\end{eqnarray}

These forms are logarithmic differentials of the functions defining
the curves: if we denote $F=(f/g)^{-1/(\alpha+1)}\xi_3\xi_2^{-1}$,
then $\gamma_2=\{F=1\}$ and $\eta_2={d\log F}$.  Similarly,
$\eta_1={d\log G}$, where $G=f^{1/(\alpha+1)}\xi_2\xi_1^{-1}$.

In what follows, we assume that 1 is a regular value of $F$ and $G$.
This can be always achieved by  a small perturbation of parameters
and is enough for the proof of Theorem \ref{uni}.b by
upper-continuity of the number of the non-degenerate critical points.

\begin{lemma}
The curves $\gamma_1$ and $\gamma_2$ are separating leaves  of the polynomial forms $\eta_1$ and $\eta_2$.
\end{lemma}

\begin{proof}
Indeed, $\gamma_2$ is a level curve of the function
$F=(f/g)^{-1/(\alpha+1)}\xi_3\xi_2^{-1},$ which is a smooth function
on $\RR^2_+$. Thus, $\gamma_2$ coincides with the boundary of the
domain $\partial\{F<1\}$. Therefore the value of $\eta_2={d(\log F)}$
on the outer normal to $\{F<1\}$ is non-negative, and is everywhere
positive since 1 is not a critical value of $F$. This means  that the
coorientations of $\gamma_2$ as the boundary of $\{F<1\}$ and as
defined by the polynomial form $\eta_2$ coincide.

Similar arguments hold for $\gamma_1$, and we conclude that
$\gamma_1$ and $\gamma_2$ are separating leaves of the
forms $\eta_1$ and $\eta_2$.
\end{proof}
%
%\begin{lemma}
%For generic values of $\alpha$ and fixed $a,b, \zeta_i$  the curve
%$\gamma_1$ does not pass through the critical points of $F$ in
%$\RR_+^2$.
%\end{lemma}
%\begin{proof}
%Indeed, this can only happen when the unique minimum of the convex
%function  $g^{-1/(\alpha+1)}\xi_2$ in $\RR_+^2$ is equal to that
%of the convex function $f^{-1/(\alpha+1)}\xi_3$, which generically
%is not the case.
%\end{proof}

This enables application of Theorem
\ref{th:Khov} to the pair $(\gamma_1,\gamma_2)$ and we obtain the
following estimate:
\begin{proposition}\label{N}
       $$N\le N_1+N_2,$$
      where $N$ is the number of points in the intersection
$\gamma_1\cap\gamma_2\cap\RR_+^2$, $N_1$ is the number of the
noncompact components of $\gamma_2$ in $\RR_+^2$ and $N_2$ is the
number of points  of intersection of $\gamma_2$ with the set
$\Gamma=\{\eta_1\wedge\eta_2=0\}$ in $\RR_+^2$.
\end{proposition}

\begin{lemma}\label{N_1} $N_1=2$.
\end{lemma}
\begin{proof}
Asymptotically the equation
$f^{-1/(\alpha+1)}\xi_3=g^{-1/(\alpha+1)}\xi_2$ has four
solutions: $g\sim\const\cdot f$ as $f\to 0$ or $\infty$,
$g\sim\const\cdot f^{-1-2\alpha}$ as $f\to\infty$, and
$f\sim\const\cdot g^{-1-2\alpha}$ as $g\to\infty$.

The number of noncompact components of $\gamma_2$ in $\RR_+^2$
equals to the half of the number of its intersection points with
boundary of a large rectangle $\{\ep_1\le f\le\ep_1^{-1}, \ep_2\le
g\le \ep_2^{-1}, 0<\ep_2\ll\ep_1\ll 1\}$. These intersection
points correspond to the above asymptotic solutions and,
therefore, their  number equals to $4$. Thus, the number $N_1$ of
unbounded components of $\gamma_2$ in $\RR_+^2$ is $2$.
\end{proof}

\subsubsection{Second application of Rolle-Khovanskii theorem}

In order to estimate $N_2$ we apply Theorem \ref{th:Khov}
again. The set $\Gamma=\{\eta_1\wedge\eta_2=0\}$ is a real
algebraic curve given by the equation $Q=0$, where
\begin{equation}
Q\,df\wedge dg=fg\xi_1\xi_2\xi_3\cdot\eta_1\wedge\eta_2
\end{equation}
is a polynomial in $(f,g)$. Applying Theorem \ref{th:Khov} again
we get the following estimate.
\begin{proposition}\label{N_2}
       $$N_2\le N_3+N_4,$$
where (as above) $N_2$  is the number of points  in
$\{\gamma_2\cap\Gamma\cap\RR_+^2\}$, $N_3$ is the number of
noncompact components of $\Gamma$ in $\RR_+^2$ and $N_4$ is the
number of  points in
$\Gamma\cap\{d{Q}\wedge{\e}=0\}\cap\RR_+^2$.
\end{proposition}

\begin{notation} For any polynomial $S$ in two variables let $\NP(S)$
denote the Newton polygon of $S$. By $'\Subset'$ we denote the
partial order by inclusion on plane polygons, namely, $'A\Subset
B'$ means that a polygon $A$ lies strictly inside a polygon $B$.
\end{notation}

\begin{lemma}
The set $\Gamma$ has no unbounded components in $\RR^2$. Moreover,
$\Gamma$ does not intersect the coordinate axes except at the
origin, which is an isolated point of $\Gamma$.
\end{lemma}

\begin{proof}
Explicit computation shows that
\begin{equation}\label{Qexact}
Q=\frac{-1-2\alpha}{(\alpha+1)^2}\xi_1\xi_2
\xi_3-fgQ_1,\quad\text{where}\quad Q_1=\xi_2'\xi_3\frac{\partial
\xi_1}{\partial f}+\xi_2\xi_3'\frac{\partial\xi_1}{\partial
g}-\xi_2'\xi_3'\xi_1.
\end{equation}

One can easily check that $\frac{\partial^3 Q_1}{\partial
f^3}=\frac{\partial^3 Q_1}{\partial g^3}=\frac{\partial^4
Q_1}{\partial f^2\partial g^2}\equiv0$ and $Q_1(0,0)=0$. This
implies that the Newton polygon of the polynomial $fgQ_1$ lies
strictly inside the Newton polygon of $\xi_1\xi_2\xi_3$:
\begin{equation*}
\NP(fgQ_1)=\{3\le p+q\le5,1\le p,q\le3\}\Subset \NP(Q)=\{2\le
p+q\le6, 0\le p,q\le4\}.
\end{equation*}
Therefore the number of unbounded components of $\Gamma$
in $\RR_+^2$ coincides with the number of unbounded components of
the zero locus of $\xi_1\xi_2\xi_3$ in $\RR_+^2$, the latter being
equal to zero.

Another proof can be obtained by parameterizing the unbounded
components of $\Gamma$ near infinity and near the axes as
$(f=Bt^{\ep_1}+... ;\; g=At^{\ep_2}+...)$. From the shape of the
Newton polygon of $Q$ one can show that $\ep_1/\ep_2$ is either
$0,1$ or $\infty$. Therefore, $B$ should be a root of $\xi_3$,
$A/B$ should be a root of $\xi_1$ or $A$ should be a root of
$\xi_2$, respectively. Since neither of them has real roots, we conclude that
$\Gamma$ has no real unbounded components.

On the coordinate axes the polynomial $Q$ equals $\xi_1\xi_2\xi_3$
and is therefore positive with the exception of the origin. The
quadratic form of $Q$ at the origin, being proportional to
$\xi_1$, is definite. Therefore the origin is an isolated zero of
$Q$ and, therefore, an isolated point of $\Gamma$.
\end{proof}

\begin{corollary} $N_3=0$.\end{corollary}

\subsubsection{Estimating the number $N_4$ of points of contact
between $\Gamma$ and $\eta_2$}

These points are zeros of the polynomial form
$$Rdf\wedge
dg=fg\xi_2\xi_3\cdot dQ\wedge\eta_2.$$
Thus we have to estimate the
number of solutions of the system $Q=R=0$ in $\RR_+^2$.

We proceed as follows. Using the Bernstein-Kushnirenko theorem we
find an upper bound on the number of solutions of $Q=R=0$ in
$(\CC^*)^2$, and then reduce it by the number of solutions known to be outside $\RR_+^2$.

The Bernstein-Kushnirenko upper bound for the number of common zeros
of polynomials $Q$ and $R$ is expressed in terms of the mixed volume
of their Newton polygons.  In fact, in computation of this mixed
volume we replace $R$ by its difference with a suitable multiple of
$Q$: this operation does not change common zeros of $Q$ and $R$, but
significantly decreases the mixed volume of their Newton polygons.

Simple  degree count shows that the Newton polygon of $R$ is given
by
$$
\NP(R)=\{2\le p+q\le10,0\le p,q\le 6\}.
$$

\begin{lemma} There exists a polynomial $q=q(f,g)$ such that the
Newton polygon of $\widetilde{R}=R-qQ$ lies strictly inside  the
Newton polygon of $R$. In other words,
$$
\NP(\widetilde R)\subseteq\{3\le p+q\le 9, 1\le p,q\le 5\}.
$$
\end{lemma}

\begin{proof}
Our goal is to prove that all monomials lying on the boundary of
$\NP(R)$ (further called boundary monomials) are equal to monomials
lying on the boundary of a Newton polygon of some multiple of $Q$. We
constantly use the fact that the boundary monomials of a product are
equal to the boundary monomials of the product of boundary monomials
of the factors.

First, let us replace $R$ by a polynomial $R_2$ with the same Newton
polygon and the same monomials on its boundary, but with simpler
definition. We have seen above that
$\NP(fgQ_1)\Subset\NP(Q)=\NP(\xi_1\xi_2\xi_3)$. Denote $R_1 df\wedge
dg=fg\xi_2\xi_3\cdot d(fgQ_1)\wedge\eta_2$. Computation of degrees
shows that
$$\NP(R_1)\subseteq\{3\le p+q\le 9, 1\le p,q\le 5\}\Subset\NP(R)=\{2\le
p+q\le 10, 0\le p,q\le 6\}.$$

Therefore one can disregard $R_1$ and consider only $R_2=R-R_1$, where
$$R_2df\wedge dg=fg\xi_2\xi_3\cdot d(Q-fgQ_1)\wedge d\eta_2=\const fg\xi_2\xi_3\cdot d(\xi_1\xi_2\xi_3)\wedge\eta_2.$$
Computing the product we see that:
\begin{eqnarray*}
R_2&=&\const\cdot
d(\xi_1\xi_2\xi_3)\wedge\left[g\xi_2\left(-\frac{\xi_3}{1+\alpha}+f\xi_3'\right)df+
f\xi_3\left(\frac{\xi_2}{1+\alpha}-g\xi_2'\right)dg\right]=
\\
&=&\const\cdot\xi_2\xi_3\left\{\vphantom{\frac 1
2}\xi_1\left[\vphantom{\RR^{\RR^\RR}_{\RR_\RR}}2\xi_2\xi_3
+f\xi_2\xi_3'+g\xi_3\xi_2'
-3(1+\alpha)fg\xi_2'\xi_3'\right]-(1+\alpha)fgQ_1\right\}.
\end{eqnarray*}
A simple computation using \ref{prop:common zero in 3dim} shows that
the Newton polygon of the first product in the figure brackets
coincides with $\NP(Q)=\NP(\xi_1\xi_2\xi_3)$. The Newton polygon of
the second product lies strictly inside of $\NP(Q)$, as was shown in
(\ref{Qexact}). Therefore it does not affect boundary monomials and
can be disregarded.

The  remaining terms sum to   $ q\xi_1\xi_2\xi_3$, where we denote
$f\xi_2\xi_3'+g\xi_3\xi_2' +2\xi_2\xi_3-3(1+\alpha)fg\xi_2'\xi_3'$ by
$q$. Up to a non-zero constant factor, its boundary monomials  are
the same as the boundary monomials of $qQ$: the polynomials $Q$ and
$\xi_1\xi_2\xi_3$ have proportional boundary monomials by
(\ref{Qexact}).

Using these facts we conclude that for
$\widetilde{R}=R-\const qQ$ one gets
$\NP(\widetilde{R})\Subset\NP(R)$.
\end{proof}

\subsubsection{Bernstein-Kushnirenko theorem}
Applying the well-known result  of \cite{Bernstein,Kushnirenko} we
know  that the number of common zeros of $Q$ and $\tilde{R}$
in $(\CC^*)^2$ does not exceed twice the mixed volume of
$\NP(Q)$ and $\NP(\widetilde{R})$.

Recall the definition of the mixed volume of two polygons. Let $A$
and $B$ be two planar convex polygons. It is a common knowledge
that the volume of their Minkowsky sum $\lambda A+\mu B$ is a
homogeneous quadratic polynomial in (positive) $\lambda$ and
$\mu$:
$$
Vol(\lambda A+\mu
B)=Vol(A)\lambda^2+2Vol(A,B)\lambda\mu+Vol(B)\mu^2.
$$
By definition the {\bf mixed volume} of two polygons $A$ and $B$
is the coefficient $Vol(A,B)$.

Setting $\lambda=\mu=1$, one gets
$$
2Vol(A,B)=Vol(A+B)-Vol(A)-Vol(B).
$$

\begin{lemma} There are at most 28
common zeros of $Q=R=0$ in $(\CC^*)^2$.\end{lemma}

\begin{proof} Simple count gives that $2Vol(\NP(Q),\NP(\widetilde
R))=28$.
       \end{proof}

         \begin{figure}[!htb]
\centerline{\hbox{\epsfysize=5cm\epsfbox{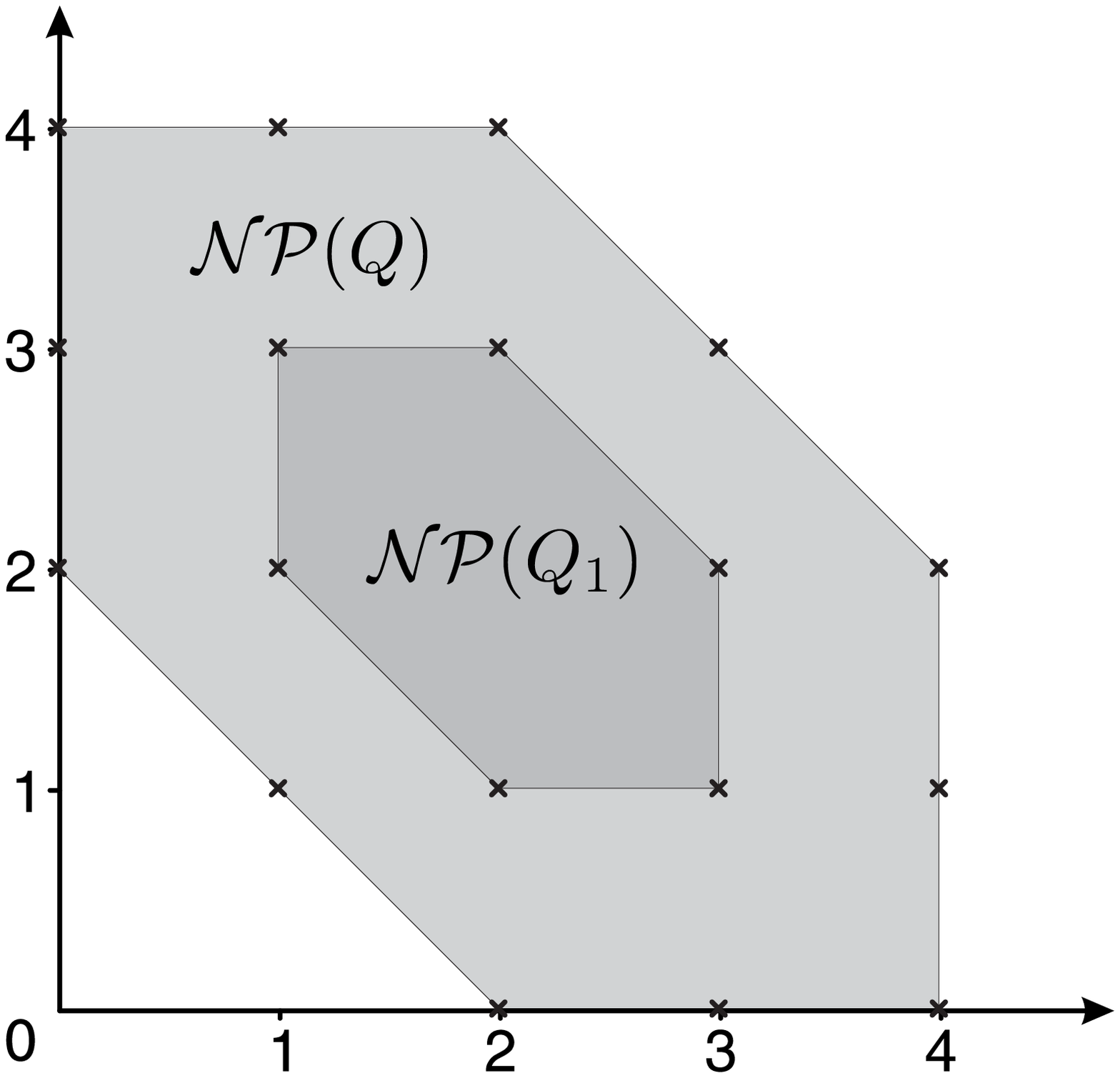}}
\hskip1cm\hbox{\epsfysize=5cm\epsfbox{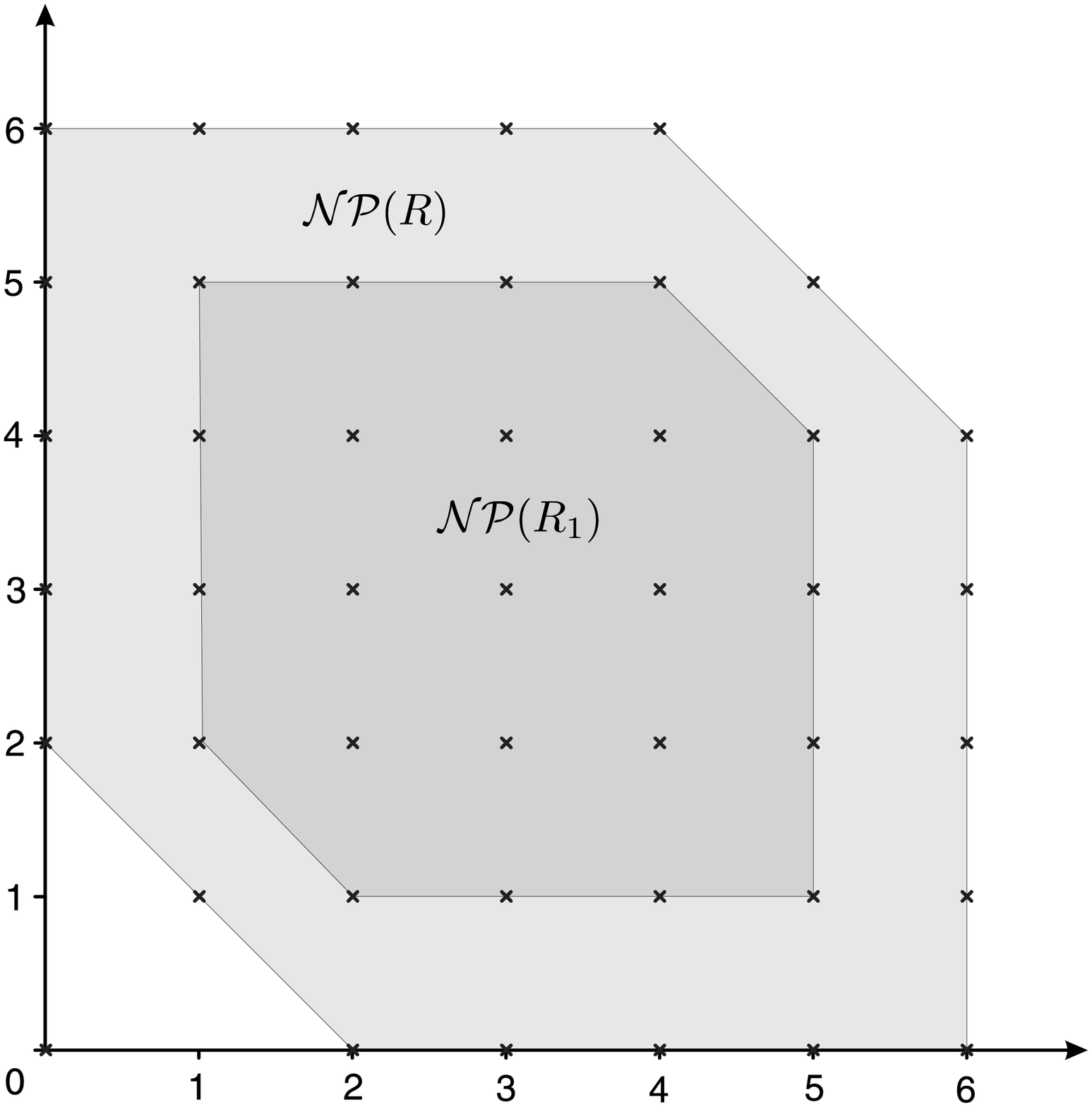}}} \caption{Relevant Newton
polygons.  }
\label{fig2}
\end{figure}

\subsubsection{Common zeros of $Q$ and $R$ outside $\RR_+^2$}

By Lemma \ref{prop:common zero in 3dim} the quadratic polynomials
$\xi_1,\xi_2$ and $\xi_3$ have two common zeros. Denote them by
$(f_1,g_1)$ and $(f_2,g_2)$. Evidently, they are not real (e.g. since
$\xi_3$ is strictly positive on $\RR^2$).
\begin{lemma}
$(f_1,g_1)$ and $(f_2,g_2)$ are solutions of the system $Q=R=0$ of
multiplicity at least 6 each.
\end{lemma}

\begin{proof}
Since $(f_1,g_1)$ and $(f_2,g_2)$ are conjugate and the system
$Q=R=0$ is real, it suffices to consider one of these points, say
$(f_1,g_1)$.

 From (\ref{Qexact}) one can immediately see that $Q(f_1,g_1)=0$.
Moreover, differentiating $Q_1$, one can see that $\frac{\partial
Q_1}{\partial f}(f_1,g_1)=\frac{\partial Q_1}{\partial
g}(f_1,g_1)=0$, so $(f_1,g_1)$ a critical point of $Q$.

Recall that
$R=dQ\wedge\left(g\xi_2(-\frac{\xi_3}{1+\alpha}+f\xi_3')df+
f\xi_3(\frac{\xi_2}{1+\alpha}-g\xi_2')dg\right)$, i.e.  $R$ is the
product of two polynomial forms each having a simple zero at
$(f_1,g_1)$. Therefore this point is necessarily a critical point
of $R$ as well.

Moreover, the 1-jet of $\eta_2$ at $(f_1,g_1)$ equals
$$
j^1\eta_2=f_1g_1\xi_2'(g_1)\xi_3'(f_1)\left[\vphantom{\|\bigotimes}(g-g_1)df-(f-f_1)dg\right],
$$
i.e. is proportional to the Euler form. Therefore, the quadratic part
of $R$ at $(f_1,g_1)$ is  proportional to the exterior product of the
differential of the quadratic part of $Q$ at $(f_1,g_1)$ and the
Euler form, so is  proportional to the quadratic form of  $Q$ at
$(f_1,g_1)$. Thus a suitable linear combination of $R$ and $Q$ has
both linear and quadratic part zero at $(f_1,g_1)$, which implies
that the multiplicity of $(f_1,g_1)$ as a solution of the system
$Q=R=0$  is at least 6.
\end{proof}

\begin{lemma} At least  six real solutions of $Q=R=0$ lie outside $\RR_+^2$.
\end{lemma}

\begin{proof}
There are exactly four points where the  form $\eta_{2}$ vanishes,
exactly one in each real quadrant. These points are evidently
solutions of the system $Q=R=0$. Consider a connected component of
the curve $\{Q=0\}$ containing such a point. It is a compact  oval
not intersecting the coordinate axes. The polynomial $R$ vanishes at
least once on this component, namely at this point. Therefore, $R$
should have at least one another zero on this oval (counting with
multiplicities), also necessarily lying in the same quadrant.
\end{proof}

\subsubsection{Final count} The number $N$ of points in the
intersection $\gamma_{1}\cap\gamma_{2}\cap \RR_+^2$ is less or
equal     $2+0+28-18=12$, where    $2$ is the number of unbounded
components of $\gamma_1$ in $\RR_+^2$; $0$ is the number of
unbounded components of $\{Q=0\}$; $28$ is the
Kushnirenko-Bernstein upper bound for the number of complex
solution of $R=Q=0$ in $(\mathbb C^{*})^2$ and $18$ is the number
of solutions of $R=Q=0$ outside $\RR_+^2$ counted with
multiplicities. Therefore, Proposition \ref{pr:12} and Theorem
\ref{uni}.b are finally proved. \qed

\subsubsection{Comments on Theorem \ref{uni}}
\label{fincom}

\text{ }\newline

{1.} Computer experiments indicate that,
except for the two solutions of the system $\xi_1=\xi_2=\xi_3=0$,
all the remaining 16 solutions of the system $R=Q=0$ can be real.
However, not all of them lie in the positive quadrant: typically
the system $R=Q=0$ has 4 solutions in each real open quadrant.
This implies that, provided that this statement about the root
configuration could be rigorously proved, the best estimate
obtainable by the  above method would be $6$, very close to
Maxwell's conjectural  bound $4$.

{2.}  An even more important observation is that there are
typically only two points of intersection of $\gamma_2$ and
$\Gamma=\{Q=0\}$ lying in $\RR_+^2$. In other words, the first
application of the Rolle-Khovanskii lemma numerically seems to be
exact: a rigorous proof that there are just two points in
$\gamma_1\cap\Gamma\cap\RR_+^2$ would imply  the
original Maxwell conjecture.

In fact, it is enough to prove a seemingly simpler statement that
the number of intersections of $\Gamma$ and
$\gamma_1=\{f^{1/(\alpha+1)}\rho_1-\rho=0\}$ lying in $\RR_+^2$ is
at most two. This seems to be easier since the equation defining
$\gamma_1$, being  quadratic polynomial in $g$, can be solved
explicitly. The resulting two solutions $g=g_{1,2}(f)$
parameterize $\gamma_1$, and the problem reduces to the question
about the number of positive zeros of a univariate algebraic function $Q(f,g_1(f))$.

{3.} The fact that the polynomial $R$ can be reduced to a smaller
polynomial $\tilde{R}$ by subtraction of a multiple of $Q$ is a
manifestation of a general yet unexplained phenomenon: tuples of
polynomials resulting from several consecutive applications of the
Rolle-Khovanskii theorem are very far from generic, and in every
specific case one can usually make a reduction similar to the
reduction of $R$ to $\tilde{R}$ above.
%\mitya

\subsection{Proof of Theorem \ref{th:asymp}.a}
 From now on we will always assume that all our charges are positive
(the case of all negative charges follows by a global sign change).

\subsubsection{$1$-dimensional case}

As a warm-up exercise we will prove Theorem \ref{th:asymp}.b in
the simplest case of $x$-axis.

The idea of the proof is to use the limit function
\begin{equation}\label{1-dim limit}
V_{\infty}(x)=\min_{i=1,\ldots,l}
((x-x_i)^2+y_i^2)=\lim_{\alpha\to\infty}V_{\alpha}^{-1/\alpha}(x),
\end{equation}
where $(x_{i},y_{i}),\;i=1,\ldots,l$ are the coordinates
of the $i$-th charge. (We assume for
simplicity that all $y_{i}\neq 0$. The general case follows by
taking the  limit.) The function $V_{\infty}(x)$ has at most $l-1$
points of non-smoothness. Denote these points by $\gamma_j$'s.

\begin{lemma} Convergence
      $V_{\infty}(x)=\lim_{\alpha\to\infty}V_{\alpha}^{-1/\alpha}(x)$
      is valid in the $C^2$-class on any closed interval free from
      $\gamma_j$'s.
\end{lemma}
\begin{proof}
We assume that on such an interval $\rho_1<(1-\e)\rho_i$, $i\ge
2$, $\e>0$. (Here $\rho_{i}=(x-x_i)^2+y_i^2$.) Therefore,
$V_{\infty}(x)=\rho_1$ on this interval. The first derivative of
$V_{\infty}(x)$ equals
\begin{eqnarray*}
(V_{\alpha}^{-1/\alpha}(x))'=-\frac 1 \alpha
V_{\alpha}^{-1/\alpha-1}(x)\left(-\alpha \sum_{i=1}^l
2\zeta_i\rho_i^{-\alpha-1}(x-x_i)\right)=\\
=2\left[\zeta_1(x-x_1)+\sum_{i=2}^l\zeta_i(\rho_i/\rho_1)^{-\alpha-1}(x-x_i)\right]
\left(\rho_1^{\alpha}V_{\alpha}(x)\right)^{-2/\alpha-1}=\\
=2[\zeta_1(x-x_1)+o(1)](\zeta_1+o(1))^{-2/\alpha-1}=2(x-x_1)+o(1)=V_\infty'(x)+o(1),
\end{eqnarray*}
where $\lim_{\al\to\infty}o(1)=0$.

Computations with the second derivative are similar, but more
cumbersome.
\end{proof}

\begin{corollary}
For $\alpha$ sufficiently large any closed interval free from
$\gamma_j$'s contains  at most one critical point of
$V_{\alpha}(x)$.
\end{corollary}

\begin{proof}
Indeed, for any sufficiently large $\alpha$  the second derivative
$(V_{\alpha}^{-1/\alpha}(x))''$, being close to $V_{\infty}''=2$,
is positive on this interval. Therefore,
$V_{\alpha}^{-1/\alpha}(x)$ is convex and can have at most one
critical point on this interval. But the critical points of
$V_{\alpha}^{-1/\alpha}(x)$ are the same as the critical points of
$V_{\alpha}(x)$.
\end{proof}

\begin{lemma}  For any sufficiently large
$\alpha$  a closed  interval containing some $\gamma_j$ and free
from $x_i$'s contains at most one critical point of
$V_{\alpha}(x)$.
\end{lemma}

\begin{proof}   Note that such an interval contains exactly one
$\gamma_{j}$ since $\gamma_{j}$'s are separated by $x_{i}$'s. The
required result follows from the fact that $V_{\alpha}(x)$ is
necessarily convex on any such interval. Indeed,
$$
(V_{\alpha}(x))''=\alpha(\alpha+1)\sum_{i=1}^l
\zeta_i\rho_i^{-\alpha-2}\left(4(x-x_i)^2-\frac
{2\rho_i}{\alpha+1}\right).
$$
Since $x-x_i\not=0$ on the interval under consideration, then $\frac
{2\rho_i}{\alpha+1}$ is necessarily smaller than $4(x-x_i)^2$ for
$\al$ large enough. Thus, $ (V_{\alpha}(x))''$ is positive (recall
that $\zeta_i>0$) and $V_{\alpha}(x)$ itself is convex.
\end{proof}

\subsubsection{\textbf{Multidimensional case}}

We start with the discussion of the critical points of the
limiting function $V_{\infty}(\bar x)$.

\subsubsection{Critical points of $V_{\infty}(\bar x)$}
\label{cr:infty}

The function $V_{\infty}(\bar x)$ is a piecewise smooth continuous
semialgebraic function. Here are the definitions of critical
points of such functions and their Morse indices adapted to our
situation.

\begin{definition} A point $\bar{x}_0$ is a {\bf critical point} of
$V_{\infty}(\bar x)$
if for any sufficiently small ball $B$ centered at $\bar{x}_0$ its
subset $ B_-=\{V_{\infty}(\bar{x})<V_{\infty}(\bar{x}_0)\}\subseteq B$
is either empty or noncontractible.

The critical point $\bar{x}_0$ is called \textbf{nondegenerate} if
$B_-$ is either empty or homologically equivalent to a sphere. In
this case the {\bf Morse index} of $\bar{x}_0$ is defined  as the
dimension of this sphere plus $1$. (By default,
$\dim(\emptyset)=-1$.)
      \end{definition}
%\mitrii{Kruto ya zagnul, a?}

\begin{lemma} Every effective Voronoi cell of the Voronoi diagram of a
generic configuration of positive charges contains
    a unique critical point of $V_{\infty}(\bar x)$. Its index equals
    the dimension of the Voronoi cell.
    \label{lm:ess}
      \end{lemma}

      \begin{proof} Indeed, take any effective Voronoi cell $S$. As above
      let $\NS(S)$ denote the set of all nearest sites of $S$. By
      definition, $S$ intersects the convex hull of $\NS(S)$. Denote
      this (unique) intersection point by $p(S)$. We claim that $p(S)$
      is the unique critical point of $V_{\infty}(\bar x)$ located on
      $S$. Indeed, the function $V_{\infty}(\bar x)$ restricted to
      $\NS(S)$ has a local maximum at $p(S)$ since any sufficiently
      small move within $\NS(S)$ brings us closer to one of the
      nearest sites. (Here we implicitly use the genericity
      assumptions on the configuration, i.e. that there are exactly
      $k+1$ nearest sites for any Voronoi cell of codimension $k$ and that
      $S$ intersects the interior of the closure of $\NS(S)$.) On the
      other hand, the restriction of $V_{\infty}(\bar x)$ to $S$
      itself has the global minimum on $S$ for similar reasons.
      \end{proof}

         \begin{figure}[!htb]
\centerline{\hbox{\epsfysize=4cm\epsfbox{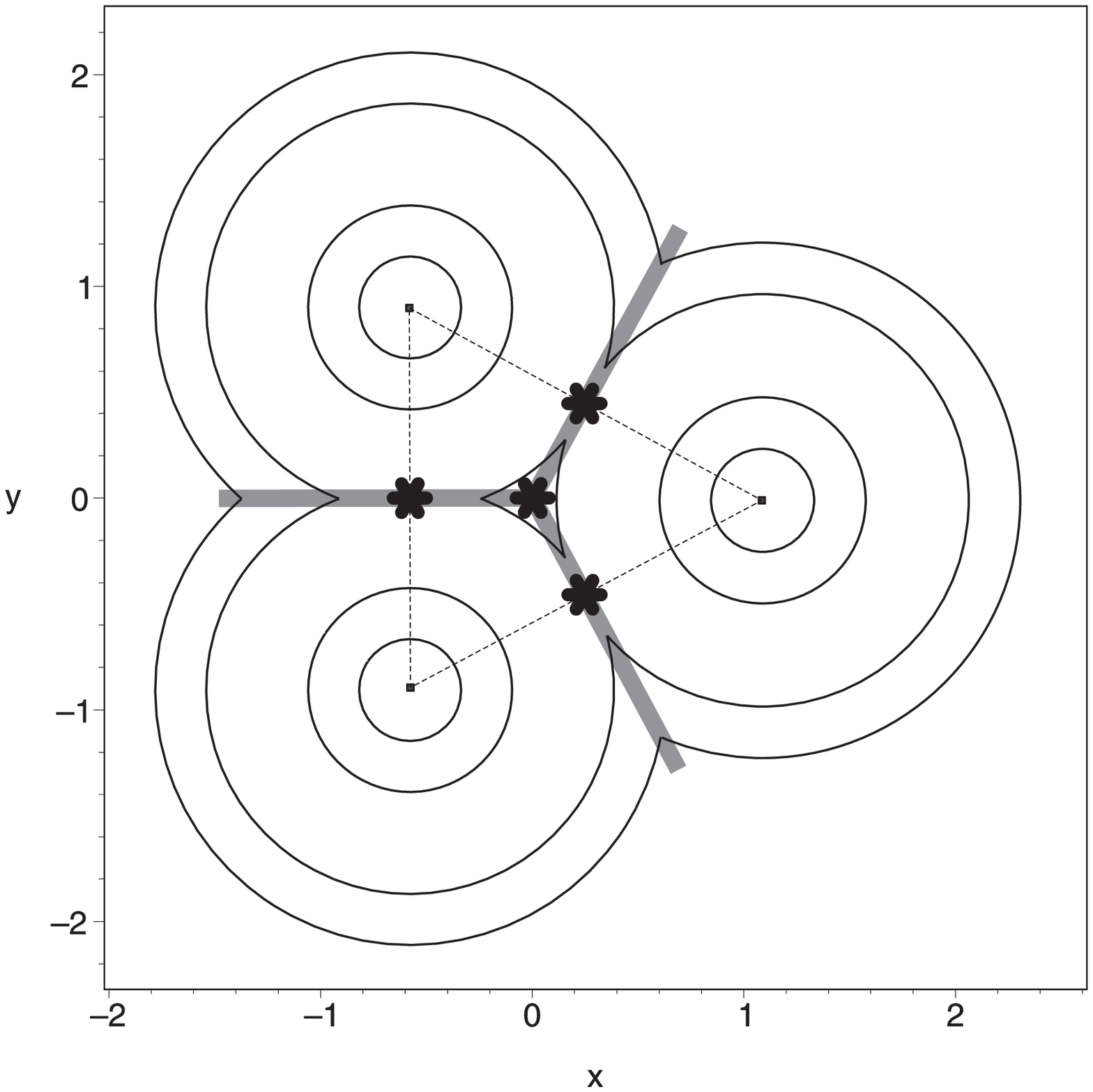}}
\hskip1cm\hbox{\epsfysize=4cm\epsfbox{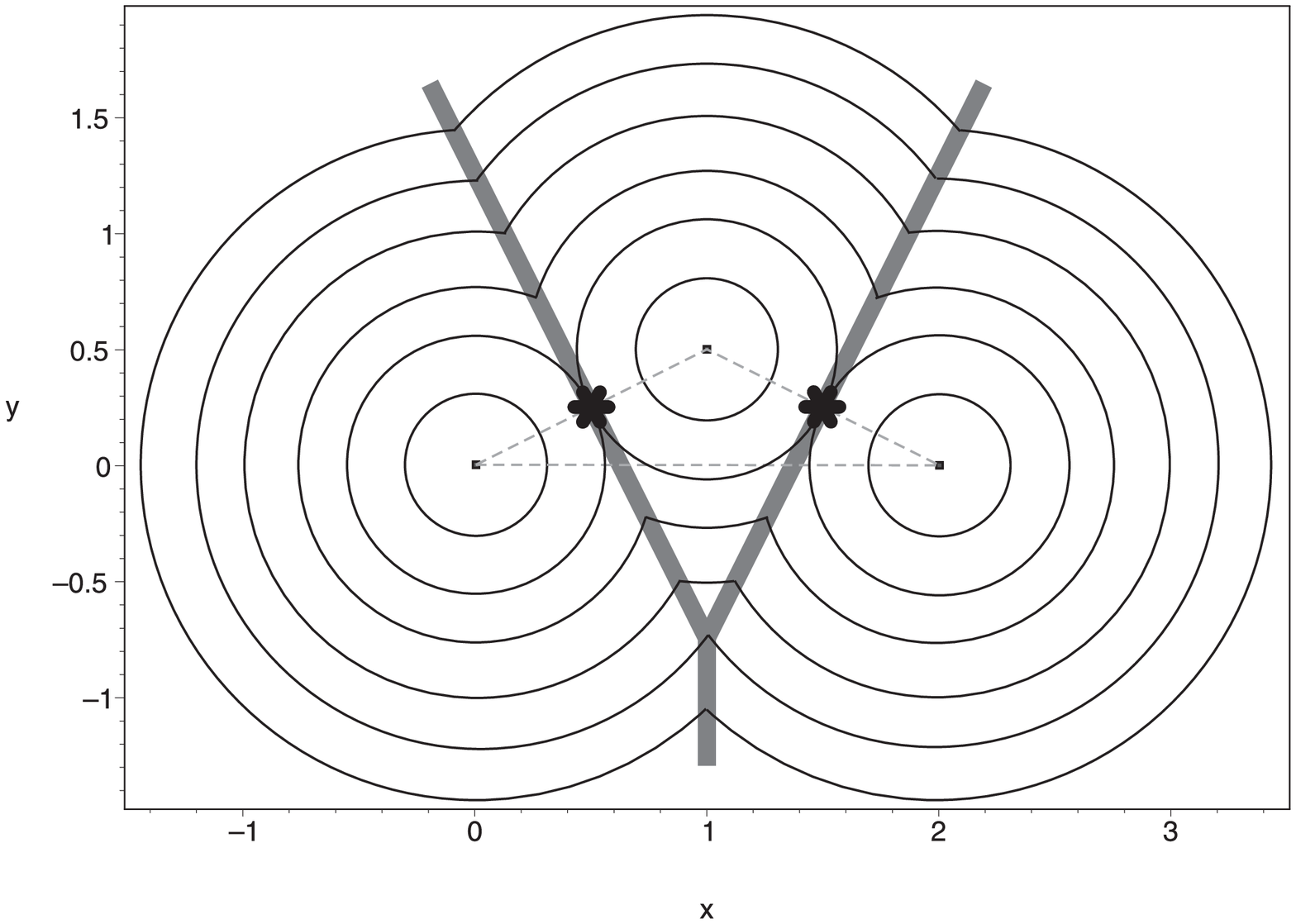}}}
\caption{Effective and ineffective $0$-dimensional Voronoi cell  of
$V_{\infty}(\bar x)$.  } \label{fig3}
\end{figure}

\noindent
{\it  Remark.}
   Figure \ref{fig3} illustrates the above Lemma \ref{lm:ess}. The left
   picture shows the function $V_{\infty}(\bar x)=\min
(\rho_{1},\rho_{2},\rho_{3})$ where
   the three points are located at $(1,0),\;$ $(\pm\frac{\sqrt{3}}2,-\frac 1 2),\;$.
   It is related to the left picture on Figure~\ref{fig1} showing the
   corresponding   potential $V_{\al}(\bar x)$ for $\al=1$. In
   this case all the Voronoi cells of the Voronoi diagram are effective and one sees
   the local maximum inside the convex hull of these points. On the right
   picture the three points are located at  $(0,0),\;(2,0),\;(1,\frac 1 2)$.
   In this case the $0$-dimensional Voronoi cell and one of $1$-dimensional
   Voronoi cells are ineffective and there is no critical point at the
   $0$-dimensional Voronoi cell.  This picture is similarly related to the
   right picture on Figure~\ref{fig1}.
%\end{remark}

\subsubsection{Proof continued}

In order to settle the multidimensional case we generalize  the
previous proof using the following idea.

\medskip\noindent\emph{ Main idea for zero-dimensional Voronoi cells\/}: Near an
effective zero-dimensional Voronoi cell of the Voronoi diagram the
union of the region where $V_{\alpha}(\bar x)$ is convex (and
therefore has at most one critical point) and the region where
$V_{\alpha}(\bar x)$ is too $\mathcal{C}^1$-close to
$V_{\infty}^{-\alpha}(\bar x)$ to have any critical points
asymptotically covers a complete neighborhood of the Voronoi cell. More exact,
 we compute asymptotics of the sizes of the above regions,
and show that the first region shrinks slower than the
second region grows. \medskip

The following expressions for the gradient and the Hessian form (i.e.,
the quadratic form defined by the matrix of the  $2$-nd partial
derivatives) of $V_{\alpha}(\bar x)$ are crucial for further
computations:
\begin{eqnarray*}
\nabla V_{\alpha}({\bar x})&=&-\alpha\sum_{i=1}^l
\zeta_i\rho_i^{-\alpha-1}(\bar{x})\nabla \rho_i(\bar{x}),
\\
\Hess V_{\alpha}(\bar x)\cdot \xi&=&\alpha(\alpha+1)\sum_{i=1}^l
\zeta_i\rho_i^{-\alpha-2}(\bar{x})\left((\nabla
\rho_i(\bar{x}),\xi)^2-\frac 2
{\alpha+1}\rho_i(\bar{x})\|\xi\|^2)\right).
\end{eqnarray*}

Here $'\, \cdot\, '$ denotes the evaluation of the quadratic form
$\Hess V_{\alpha}(\bar x)$ at $\xi$.

We start with the case of zero-dimensional Voronoi cells. The general
case will be a treated as a direct product of the zero-dimensional
case in the direction transversal to the Voronoi cell and the
full-dimensional case along the Voronoi cell.

\subsubsection{Zero-dimensional Voronoi cells of  Voronoi diagram.}

Let $S$ be a zero-dimensional Voronoi cell of a Voronoi diagram of a
generic configuration. We can assume that
$\rho_1(S)=...=\rho_{n+1}(S)<\rho_i(S)$ for $i>n+1$. Set
$$
\phi(\bar
x)=\log\left(\frac{\max_{i=1,...,n+1}\rho_i(\bar{x})}{\min_{i=1,...,n+1}\rho_i(\bar{x})}\right).
$$
Note that $\phi(\bar x)$ is everywhere positive except at the
origin,  and is equivalent to the Euclidean distance to $S$ in a
sufficiently small neighborhood of $S$.

\begin{lemma}\label{lem:choice of U}
There exists $\delta>0$ so small that in the $\delta$-neighborhood
$\mathcal{U}$ of $S$ the following conditions hold:
\begin{enumerate}

\item
There exists a number $\epsilon>0$ such that for any
$\bar{x}\in\mathcal{U}$
$$\min_{i=n+1,n+2,\ldots,l}
\rho_i(\bar{x})>e^{2\epsilon}\min_{i=1,2,\ldots,n+1}\rho_i(\bar{x})>e^{\epsilon}
\max_{i=1,\ldots,n+1}\rho_i(\bar{x}).$$ In particular,
$\phi(\bar x)<\epsilon$ in $\mathcal{U}$.
\item The absolute value of all ratios $c_k(\bar x)/c_l(\bar x)$ in the
unique linear dependence $\sum_{i=1}^{n+1} c_i(\bar x)\nabla
\rho_i(\bar{x})=0$  is bounded by some constant $\Upsilon>0$ (by the
genericity of configuration none of $c_k(S)$'s vanishes on $S$ and,
therefore, in some neighborhood of $S$ as well).

\item If the cell $S$ is not effective, then the
closure of $\mathcal U$ can be separated from the convex hull of
$\NS(S)$ by a hyperplane.\end{enumerate}
\end{lemma}

We prove that for $\alpha$ sufficiently large the domain
$\mathcal{U}$ is the union of two subdomains $\mathcal U=U_{1}\cup
U_{2}$ such that $V_{\alpha}(\bar x)$ is convex in $U_{1}$, and
$\nabla V_{\alpha}(\bar x)\not=0$ in $U_{2}$.

In what follows we denote by $C_k$ and $\kappa_k$ positive constants
independent of  $\alpha$ but dependent on the configuration and the
choice of $\mathcal{U}$.

\begin{lemma}\label{lem1:dim=0} There exists a constant $\kappa_1$
independent of $\al$ such that for $\al$ sufficiently large the
function $V_{\alpha}(\bar x)$ has no critical points in the domain
defined by $\{\phi(\bar
x)>\frac{\kappa_1}{\alpha+1}\}\cap\mathcal{U}.$
\end{lemma}
\begin{proof}
First, consider the case  $l=n+1$.

The condition $\nabla V_{\alpha}(\bar x)=0$ implies that
$$\frac{\zeta_i}{\zeta_j}\left(\frac{\rho_i}{\rho_j}\right)^{-\alpha-1}=\frac{c_j}{c_j}\le
\Upsilon,$$ and, taking the logarithm of the both sides, we arrive at
$\phi(\bar x)<\frac{\log
\Upsilon+\log\max_{i,j}(c_i/c_j)}{\alpha+1}$. Thus one can
take $\kappa_1={\log \Upsilon+\log\max_{i,j}(c_i/c_j)}$ in this case.

The case $l>n+1$ differs by exponentially small terms. Namely,
suppose that $\rho_1(\bar x)=\max\rho_i(\bar x)$. Then
$$
0=\rho_1^{\alpha+1}\nabla V_{\alpha}(\bar x)=(\zeta_1\nabla
\rho_1-\xi) +
\sum_{i=2}^{n+1}\zeta_i\left(\frac{\rho_i}{\rho_1}\right)^{-\alpha-1}\nabla
\rho_i,
$$
where $\xi=\sum_{i=n+2}^{l}\zeta_i(\rho_i/\rho_1)^{-\alpha-1}\nabla
\rho_i$. One can easily see that $\|\xi\|\le C_1
e^{-\epsilon(\alpha+1)}$. Therefore, since $\nabla\rho_2(x),...,
\nabla\rho_{n+1}(x)$ are linearly  independent in $\mathcal U$,
$$|\zeta_i(\rho_i/\rho_1)^{-\alpha-1}-\zeta_1c_i/c_1|\le C_2
e^{-\epsilon(\alpha+1)}=o(1),$$  and we get the required estimate.
\end{proof}

\begin{lemma}\label{lem2:dim=0}
There exists a constant $\kappa_2$ independent of $\al$ such that for
all sufficiently large $\al$ the function $V_{\alpha}(\bar x)$ is
convex in the domain $\{\phi(\bar
x)<\frac{\log(\alpha+1)-\kappa_2}{\alpha+2}\}$.
\end{lemma}

\begin{proof}
Again, start with the case $l=n+1$. The gradients $\nabla \rho_i,
i=1,...,n+1,$ span the whole $\RR^n$. Thus, the quadratic form
$\sum_{i=1}^{n+1}\zeta_i(\nabla \rho_i\cdot\xi)^2\ge C_3\|\xi\|^2>0$
is positive definite. Therefore, one gets
\begin{eqnarray*}
\frac{1}{\alpha(\alpha+1)} {\Hess V_{\alpha}(\bar x)}\cdot \xi=
\sum_{i=1}^{n+1}\zeta_i\rho_i^{-\alpha-2} (\nabla \rho_i,\xi)^2-\frac
2 {\alpha+1}\sum_{i=1}^{n+1}\zeta_i\rho_i^{-\alpha-1} \|\xi\|^2\ge
\\
\ge(\max_{i=1,...,n+1} \rho_i)^{-\alpha-2} \left(
\sum_{i=1}^{n+1}\zeta_i(\nabla \rho_i,\xi)^2\right)-\frac
{2(n+1)(\min_{i=1,...,n+1}
\rho_i)^{-\alpha-1}\max_{i=1,...,n+1}\zeta_i} {\alpha+1}\|\xi\|^2\ge
\\
\ge(\min_{i=1,...,n+1} \rho_i)^{-\alpha-2}\left( C_3\cdot (e^{\phi(\bar
x)})^{-\alpha-2}-\frac {C_4} {\alpha+1}\right)\|\xi\|^2.
\end{eqnarray*}
   The last form is positive
definite if
\begin{equation}\label{eq:lem2 dim=0}
e^{-(\alpha+2)\phi(\bar x)}>\frac {C_5} {\alpha+1}\text{  or,
equivalently, }\phi(\bar
x)<\frac{\log(\alpha+1)-\kappa_2}{\alpha+2}.\end{equation}

The  case $l>n+1$ differs by an exponentially small term,
namely by the term
$$\left|\sum_{i=n+2}^{l}\zeta_i\rho_i^{-\alpha-2}\left[ (\nabla
\rho_i, \xi)^2-\frac {2 \rho_i} {\alpha+1} \|\xi\|^2\right]\right|\le
C_6(\min_{i=1,...,n+1} \rho_i)^{-\alpha-2}
e^{-\epsilon(\alpha+2)}\|\xi\|^.$$
Therefore, instead of (\ref{eq:lem2
dim=0}) we get that  $\Hess V$ is positive definite provided
$$
e^{-(\alpha+2)\phi(\bar x)}>\frac {C_5} {\alpha+1}+C_6e^{-\epsilon(\alpha+2)},
$$
which gives the same estimate with a different constant.
    \end{proof}

\begin{lemma}\label{lem3:dim=0}
$V_{\alpha}(\bar x)$ has at most one critical point in
$\mathcal{U}$. If the cell under consideration is effective
then the critical point exists and is a local minimum. If
the cell under consideration is not effective then
there is no critical point in $\mathcal U$.
\end{lemma}
\begin{proof}
Indeed, in the above notation for sufficiently large $\al$ one
has
$$\frac{\log(\alpha+1)-\kappa_2}{\alpha+2}>\frac{\kappa_1}{\alpha+1}.$$
Thus $\mathcal U$ is covered by two domains, $\{\phi(\bar
x)>\frac{\kappa_1}{\alpha+1}\}$ and $\{\phi(\bar
x)\le\frac{\log(\alpha+1)-\kappa_2}{\alpha+2}\}$. By Lemma \ref{lem1:dim=0}
  $V_{\alpha}(\bar x)$ has no critical points in the first domain.
By Lemma \ref{lem2:dim=0}
 $V_{\alpha}(\bar x)$ is convex and has at most one critical
point in the second domain.

In the case when the considered $0$-dimensional Voronoi cell is
effective $V_{\alpha}(\bar x)$ actually has a local minimum located
close to that Voronoi cell: the function
$V_{\alpha}^{-1/\alpha}(\bar x)$, being $\mathcal{C}^0$-close to
$V_{\infty}$, has a local minimum inside $\mathcal U$.

The last statement is a particular case of the Lemma \ref{cor:no
crit pts for noneff} below.
\end{proof}

Taken together, this proves that for $\al$ sufficiently large to each
effective zero-dimensional Voronoi cells of a generic configuration
of points corresponds exactly one minimum of $V_{\al}$.

\subsubsection{Case of arbitrary codimension}
Let $S$ be any Voronoi cell of codimension $k$ of the Voronoi
diagram.  We prove that for a generic configuration of positive
charges and any compact $K\subset S$ lying inside  $S$  there exists
a sufficiently small neighborhood $\mathcal{U}_K$ independent of
$\alpha$ containing at most one critical point of $V_{\alpha}(\bar
x)$. Moreover, this critical point exists if and only if the cell is
effective, and its Morse index is equal to $n-k$.

Denote by $L$ the affine subspace spanned by $S$. Recall that the
first genericity assumption means that  there exist exactly $k+1$
charges $\zeta_1,...,\zeta_{k+1}$ closest to $S$.  Denote the affine
subspace orthogonal to $L$ spanned by these charges by $M$.

\begin{lemma}
$\dim M=k$.
\end{lemma}
\begin{proof}
Indeed, a small shift of any point of $S$ in \emph{any} direction
orthogonal to  $M$ produces a point with the same set of closest
charges: distances to charges not in $\NS(S)$ will still remain
bigger than the distances to the charges in $\NS(S)$, and the latter
distances will remain equal. Therefore the shifted point still lies
in $S$,  so the dimension of $S$ is at least $\codim M$, i.e., $\dim
M\ge k$. The opposite inequality is evident since $\NS(S)$ contains
$k+1$ points.
\end{proof}

If the Voronoi cell $S$ intersects the convex hull of $\NS(S)$ then
the second genericity assumption means that any $k$ of charges in
$\NS(S)$ do not lie on a hyperplane in $M$ passing through the point
$L\cap M$.

Choosing an appropriate coordinate system we may assume that $L$ and
$M$ intersect at the origin, i.e. are orthogonal complements of each
other. Denote by $\bar x_L$ and $\bar x_M$ orthogonal projections of
a vector $\bar x$ to linear subspaces $L$ and $M$ resp. i.e. $\bar
x=\bar x_M+\bar x_L$. Finally, denote the distances from $\bar x$ to
the charges $\zeta_1,...,\zeta_{k+1}$ in  $\NS(S)$ by
$\rho_1,...,\rho_{k+1}$ resp.

Let $K$ be a compact subset of $S$.

\begin{lemma}\label{lem:choice of Uk}
There exists   $\delta>0$
so small that in the $\delta$-neighborhood $\mathcal{U}_K\subset\RR^n$
of $K$ the following conditions hold:
\begin{enumerate}
\item
$\exists\,0<\epsilon \ll 1$ such that for any $\bar
x\in\mathcal{U}_K$ one has $$\min_{i=k+2,\ldots,l}
\rho_i>e^{2\epsilon} \min_{i=1,\ldots,k+1}\rho_i>e^{\epsilon}
\max_{i=1,\ldots,k+1}\rho_i.$$ This is possible since
$K$ is a compact
subset of an open Voronoi cell $S$, and is therefore located on some
positive distance from other Voronoi cells.

\item  The absolute value of all ratios $c_k(\bar x)/c_l(\bar x)$ in the
unique linear dependence $\sum_{j=1}^{k+1} c_k(\bar x)\nabla_M
\rho_j=0$  is bounded from above by some constant $\Upsilon$, where $\nabla_M
\rho_j$ denotes the orthogonal projection of the gradient $\nabla
\rho_j$ to $M$. Note that the tuple of $c_k(\bar x)$ is, up to
proportionality,  constant on $S$,  and none of $c_k$ vanishes due
to the second genericity assumption.

\item If the cell $S$ is not effective, then the closure of $\mathcal U$ can
be separated from the convex hull of $\NS(S)$ by a hyperplane.
\end{enumerate}
\end{lemma}

As before, introduce the function
$$
\phi_M(\bar
x)=\log\left(\frac{\max_{i=1,\ldots,k+1}\rho_i}{\min_{i=1,\ldots,k+1}\rho_i}\right).
$$
This function is equivalent to $\|\bar x_M\|$ near the origin:
\begin{equation}\label{eqn:phi and M_norm}
C_M^{-1}\phi_m(\bar x)\le \|\bar x_M\|\le C_M\phi_M(\bar x)
\end{equation}
for some $C_M>0$ and all $x\in\mathcal U$.

\begin{lemma}\label{lem1:dim=k}

a) For a certain positive $\kappa_3$ the function $V_{\alpha}(\bar
x)$ has no critical points in the domain given by $\mathcal{U}_K\cap\{\|\bar
x_L\|>\kappa_3\cdot e^{-\epsilon(\alpha+1)}\}$.

b) For a certain positive $\kappa_4$ the function $V_{\alpha}(\bar
x)$ has no critical points in the domain given by
$\mathcal{U}_K\cap\{\phi_M(\bar x)>\frac{\kappa_4}{\alpha+1}\}$.

\end{lemma}

\begin{proof}
a) The idea is that outside $L$ the gradients of $\rho_i$'s,
$i=1,...,k+1$, are all directed away from $L$. The  contribution of
the remaining $\rho_i$'s, being exponentially small, is negligible  outside an
exponentially small neighborhood.

We calculate the directional derivative of $V_{\alpha}(\bar x)$
in the direction $\bar x_L$  at a point $\bar x\in\mathcal{U}_K$.

$$
-\frac 1 \alpha\frac{\partial V_{\alpha}(\bar x)}{\partial \bar
x_L}(\bar x)=\sum_{i=1}^{k+1}\zeta_i\rho_i^{-\alpha-1}\frac{\partial
\rho_i}{\partial \bar x_L}(\bar
x)+\sum_{i=k+2}^{l}\zeta_i\rho_i^{-\alpha-1}\frac{\partial
\rho_i}{\partial \bar x_L}(\bar x).
$$
Since $\frac{\partial \rho_i}{\partial \bar x_L}(\bar x)=2\|\bar
x_L\|^2$ for $i=1,...,k+1$, we conclude that the absolute value of
the first term is at least
$C_9(\max_{i=1,\ldots,k+1}\rho_i)^{-\alpha-1}\|\bar  x_L\|^2$.  The
absolute value of the second term is at most $$C_{10}
(\min_{i=k+2,\ldots,l}\rho_i)^{-\alpha-1}\|\bar
x_L\|<C_{10}(\max_{i=1,\ldots,k+1}\rho_i)^{-\alpha-1}e^{-\epsilon(\alpha+1)}\|\bar
x_L\|,$$ and the estimate follows.

b) We essentially repeat the computations of Lemma
\ref{lem1:dim=0}.
\begin{eqnarray*}
-\frac 1 \alpha\nabla_M V_{\alpha}(\bar
x)=\sum_{i=1}^{k+1}\zeta_i\rho_i^{-\alpha-1}\nabla_M \rho_i(\bar
x)+\sum_{j=k+2}^{l}\zeta_i\rho_j^{-\alpha-1}\nabla_M
\rho_j(\bar x)=\\
=(\max_{j=1,\ldots,k+1}\rho_j)^{-\alpha-1}\left[\sum_{i=1}^{k+1}\zeta_i\left(\frac{\rho_i}
{\max_{j=1,\ldots,k+1}\rho_j}\right)^{-\alpha-1}\nabla_M \rho_i(\bar
x) + O(e^{-\epsilon(\alpha+1)})\right].
\end{eqnarray*}

So $\nabla_M V_{\alpha}(\bar x)=0$ implies, as in Lemma
\ref{lem1:dim=0},  that the functions
$\left(\frac{\rho_i}{\max_{j=1,\ldots,k+1}\rho_j}\right)^{-\alpha-1}$
are bounded, which gives:\; $\phi_M(\bar
x)<\frac{\kappa_4}{\alpha+1}$.

\end{proof}

\begin{lemma}\label{cor:no crit pts for noneff}
If the cell $S$ is not effective then for $\alpha$ large enough the
function $V_{\alpha}(\bar x)$ has no critical points in
$\mathcal{U}_K$.
\end{lemma}

\begin{proof}
Let $\mathfrak{n}$ be a direction normal to the hyperplane
separating $\mathcal{U}_K$ from the convex hull of $\NS(S)$. Then
$\zeta_i(\nabla\rho_i,\mathfrak{n})$ are all of the same sign in
$\mathcal U$ (say, negative), and of absolute value greater than
some positive constant $C_7$. Therefore
\begin{eqnarray*} |(\nabla
V_{\alpha},\mathfrak{n})|&=&\alpha|\sum_{i=1}^l
\zeta_i\rho_i^{-\alpha-1}(\nabla\rho_i,\mathfrak{n})|\ge
\\
&\ge&
\alpha\left[ C_7(k+1)(\max_{i\le k+1}\rho_i)^{-\alpha-1}-C_8(\min_{i>k+1}\rho_i)^{-\alpha-1}\right]\ge
\\
&\ge&\alpha (\max_{i=1,...,k+1}\rho_i)^{-\alpha-1}\left((k+1)C_7-C_8e^{-\epsilon(\alpha+1)}\right)>0
\end{eqnarray*}
for $\alpha$ large enough.
\end{proof}

\medskip
From now on we suppose that  $K$ and the convex hull of $\NS(S)$
intersect at the origin, and consider the domain
\begin{equation}\label{def Ualpha}
\mathcal{U}_{\alpha}=\left\{ \bar x_L<\kappa_3\cdot
e^{-\epsilon(\alpha+1)},\|\bar x_M\|<\frac{C_M\kappa_4}{\alpha+1}
\right\} \subset\mathcal{U}_K.
\end{equation}
Union of this domain and  the domain described in Lemma
\ref{lem1:dim=k} covers $\mathcal U$ by (\ref{eqn:phi and M_norm}).
Moreover, $\phi_M(\bar x)\le\frac{C_M^2\kappa_4}{\alpha+1}$ in
$\mathcal U$.

Our next goal is to study the quadratic form $\Hess V_{\alpha}(\bar
x) $ in the domain $\mathcal{U}_{\alpha}$.

\begin{lemma}\label{lem2:dim=k}
Let $V_{k,\alpha}(\bar x)=\sum_{i=1}^{k+1}\zeta_i\rho_i^{-\alpha}$.
The following holds

a)$\Hess V_{\alpha}(\bar x)-\Hess V_{k,\alpha}(\bar
x)=(\max_{i=1,\ldots,k+1}\rho_i)^{-\alpha-2}O(e^{-\epsilon\alpha})$
for $\bar x\in\mathcal{U}_{\alpha}$ and $\alpha\to\infty$.

b) There exist two quadratic forms $A(x)$ and $B(x)$ on $M$ and $L$
resp. such that $\Hess V_{k,\alpha} (\bar x)-A(\bar x)\oplus B(\bar
x)=(\max_{i=1,\ldots,k+1}\rho_j)^{-\alpha-2}O(e^{-\epsilon\alpha})$.

There exist some positive constants $\kappa_5, \kappa_6$ such that
the  form $A$ is positive definite and bounded from below by a
$\kappa_5\cdot\alpha^2(\max_{i=1,\ldots,k+1}\rho_i)^{-\alpha-2}$, and
the form $B$ is negative definite and  bounded from above by
$-\kappa_6\cdot\alpha(\max_{i=1,\ldots,k+1}\rho_i)^{-\alpha-2}$.
\end{lemma}

\begin{proof}
a) We have to estimate from above the contribution of the distant
charges.

\begin{eqnarray*}
&|\Hess V_{\alpha} (\bar x)\cdot \xi-\Hess V_{k,\alpha}(\bar x)\cdot
\xi|\le\alpha(\alpha+1)\sum_{i=k+2}^{l}
\zeta_i\rho_i^{-\alpha-2}((\nabla \rho_i,\xi)^2-\frac
2{\alpha+1}\rho_i\|\xi\|^2)\le\\
&\le
C\dot(\max_{i=1,\ldots,k+1}\rho_i)^{-\alpha-2}e^{-\epsilon(\alpha+2)}\|\xi\|^2.\end{eqnarray*}

b) For $i=1,...,k+1$ the charges $\zeta_i$ are in $M$. Therefore we have
\begin{eqnarray*}
(\nabla \rho_i(\bar x),\xi)^2=(\nabla_M \rho_i(\bar
x),\xi_M)^2+2(\nabla_M \rho_i(\bar x),\xi_M)(\nabla_L \rho_i(\bar
x),\xi_L)+(\nabla_L \rho_i(\bar x),\xi_L)^2=
\\
=(\nabla_M \rho_i(\bar
x),\xi_M)^2+O(e^{-\epsilon\alpha})\|\xi\|^2.
\end{eqnarray*}
Here we used $\nabla_L \rho_j(\bar x)=2\bar x_L$, and $\|\bar
x_L\|\le \kappa_3 e^{-\epsilon\alpha}$ in $\mathcal{U}_{\alpha}$.

Therefore,
\begin{eqnarray*}
\Hess V_{k,\alpha}(\bar x)\cdot \xi
=\alpha(\alpha+1)\sum_{i=1}^{k+1}\zeta_i
\rho_i^{-\alpha-2}\left[(\nabla \rho_i,\xi)^2-\frac
2{\alpha+1}\rho_i\|\xi\|^2\right]=
\\
=\alpha(\alpha+1)\sum_{i=1}^{k+1}\zeta_i \rho_i^{-\alpha-2}(\nabla_M
\rho_i,\xi_M)^2- 2\alpha\sum_{i=1}^{k+1}\zeta_i
\rho_i^{-\alpha-1}\|\xi\|^2 +O(e^{-\epsilon\alpha})\sum_{i=1}^{k+1}
\rho_i^{-\alpha-2}\|\xi\|^2.
\end{eqnarray*}
Due to linear independence of $\nabla_M \rho_j$'s in
$\mathcal{U}_\al$   we get, exactly as in Lemma \ref{lem2:dim=0},
$$\sum_{i=1}^{l}\zeta_i(\nabla_M \rho_i,\xi_M)^2>C_9\|\xi_M\|^2$$  for some
$C_9>0$. Therefore, one can estimate the first term from below as
\begin{eqnarray*}
\alpha(\alpha+1)\sum_{i=1}^{k+1}\zeta_i \rho_i^{-\alpha-2}(\nabla
\rho_j,\xi_M)^2\ge\alpha(\alpha+1)C_9(\max_{i=1,\ldots,k+1}\rho_i)^{-\alpha-2}\|\xi_M\|^2.
\end{eqnarray*}
The second term  can be estimated  using
\begin{eqnarray*}
(\min \zeta_i)(\max_{j=1,\ldots,k+1}\rho_j)^{-\alpha-1} \le \frac
1{k+1}\sum_{j=1,\ldots,k+1}\zeta_i\rho_j^{-\alpha-1}\le
(\max\zeta_i)(\max_{j=1,\ldots,k+1}\rho_j)^{-\alpha-1}e^{(\alpha+1)\phi_M(\bar
x)}.
\end{eqnarray*}
 The expression $e^{(\alpha+1)\phi_M(\bar
x)}$ is bounded in $\mathcal{U}_{\alpha}$ by a constant
$e^{C_M^2\kappa_4}$ independent of $\alpha$. Therefore
the restriction of $\Hess V_{k,\alpha}(\bar x)\cdot \xi$   to
$M$ has the lower bound:
$$
\alpha(\alpha+1)(\max_{i=1,\ldots,k+1}\rho_i)^{-\alpha-2}\left(C_9-\frac{2(k+1)e^{C_M^2\kappa_4}\max_{i,\mathcal
U}\zeta_i\rho_i}{\alpha+1}-O(e^{-\epsilon\alpha})\right)\|\xi_M\|^2.
$$

On the other hand, the restriction of $\Hess
V_{k,\alpha}(\bar x)\cdot \xi$ to $L$  is negative definite and has the
upper bound:
$$
-2\alpha(k+1)\min\zeta_i(\max_{i=1,\ldots,k+1}\rho_i)^{-\alpha-1}\|\xi_L\|^2\left(1-O(e^{-\epsilon\alpha})\right).
$$
\end{proof}

\begin{corollary}\label{cor:hess}
For large enough $\alpha$ and any $\bar x\in\mathcal{U}_{\alpha}$
the signature of the quadratic form $\Hess V_{\alpha}(\bar x)$ is
$(k,n-k)$.
\end{corollary}

\begin{lemma}
For sufficiently large $\al$ there is at most one critical point of
$V_{\alpha}(\bar x)$ in $\mathcal{U}_K$, and its Morse index is
equal to $n-k$.
\end{lemma}
\begin{proof}
As was proved in Lemma \ref{lem1:dim=k}, there are no critical
points of $V_{\alpha}$ in
$\mathcal{U}_K\setminus\mathcal{U}_{\alpha}$ for $\alpha$ large
enough. So it is enough to prove that the mapping $dV_{\alpha}$ is
one-to-one in $\mathcal U_{\alpha}$.

Take any  segment $I=\{a_t=a+t\xi,0\le t\le
1\}\subset\mathcal{U}_{\alpha}$. Let $\pi(t)$ be the projection of
the point $dV_{\alpha}(a_t)$ on the direction $\bar{\xi}=\xi_M-\xi_L$
(where $\xi=\xi_m+\xi_L$). We claim that $\pi(t)$ is a monotonous
function of $t$, and, therefore, its values at $a_0$ and $a_1$ cannot
coincide. Indeed, using Lemma \ref{lem2:dim=k} one can estimate
$\pi'(t)$ from below as
\begin{eqnarray*}
\pi'(t)=\left(\bar{\xi},\frac{\partial}{\partial t}dV_{\alpha}(a_t)\right)=\Hess V_{\alpha}(\bar{\xi},\xi)\ge \\
\ge(\max_{i=1,\ldots,k+1}\rho_i)^{-\alpha-2}\left[\kappa_5\cdot\alpha^2\|\xi_M\|^2+
\kappa_6\cdot\alpha\|\xi_L\|^2
+O(e^{-\epsilon\alpha})\|\xi\|^2\vphantom{\frac{\RR^\RR_\RR}2}\right]>0
\end{eqnarray*}
for $\alpha$ sufficiently large.

This, by convexity of $\mathcal{U}_{\alpha}$, immediately
implies the claim of the Lemma: assuming that there are two critical
points and joining them by a segment we get a
contradiction.\end{proof}

We proved in Lemma \ref{cor:no crit pts for noneff} that
non-effective cells do not create critical points of $V_{\alpha}$
for $\alpha$ large enough. To finish the proof of the one-to-one
correspondence between effective cells and critical points of
$V_{\alpha}$ we have to show that if $K$ contains a critical point
of the function $V_{\infty}$  then $\mathcal{U}_K$ does contain a
critical point of $V_{\alpha}$.

\begin{lemma}\label{lem:critpts near effective}
Assume that $K$ contains a critical point of $V_{\infty}$. Then
$\mathcal{U}_K$ contains a critical point of $V_{\alpha}$ for
$\alpha$ sufficiently large.
\end{lemma}

The idea of the proof is that the smooth function
$V_{\alpha}^{-\frac 1 {\alpha}}$ is arbitrarily
$\mathcal{C}^0$-close to $V_{\infty}$ for large $\alpha$. The
topology of the sets $X_c=\{V_{\infty}\le c\}\cap{\mathcal{U}_K}$
changes as $c$ passes the critical value implying the change of the topology of the
sets $Y_c=\{V_{\alpha}^{-\frac 1 {\alpha}}\le
c\}\cap{\mathcal{U}_K}$ for sufficiently large $\alpha$ which in its
turn implies presence of the
critical points of $V_{\alpha}^{-\frac 1 {\alpha}}$, the latter
coinciding with the critical points of $V_{\alpha}$.

Let $c_0$ be the critical value of $V_{\infty}$ at the point $c\in
K$, and let $c_1$, $c_2$ be some regular values,
$\min_{\mathcal{U}_K} V_{\infty}<c_1<c_0<c_2<\max_{\mathcal{U}_K}
V_{\infty}$. We assume that $c_1,c_2$ are so close to $c_0$ that the
interval $[c_1,c_2]$ contains no critical values of $V_{\infty}$
restricted to the boundary of $\mathcal{U}_K$.

Let $\delta\ll (c_0-c_1)/2$.  For $\alpha$ large enough the function
$V_{\alpha}^{-\frac 1 {\alpha}}$ is at least $\delta/2$-close to
$V_{\infty}$. Thus, we have
$$X_{c_1}\subset Y_{c_1+\delta/2}
\subset X_{c_1+\delta}\subset Y_{c_1+\frac 3 2\delta}.$$

These inclusions induce homomorphisms in homology groups:
\begin{equation}\label{eqn:homos}
H_*(X_{c_1})\to H_*(Y_{c_1+\delta/2})\to H_*(X_{c_1+\delta})\to
H_*(Y_{c_1+\frac 3 2\delta}).
\end{equation}
The composition of the first two homomorphisms is a homomorphism
induced by the inclusion $X_{c_1}\subset X_{c_1+\delta}$, which is an
isomorphism. Therefore the middle homomorphism  in (\ref{eqn:homos})
is surjective.

Similarly, the composition of the last two homomorphisms is a
homomorphism induced by the inclusion $Y_{c_1+\delta/2}\subset
Y_{c_1+\frac 3 2\delta}$, which,  assuming that  $V_{\alpha}^{-\frac
1 {\alpha}}$ has no critical points, is an isomorphism as well.
Therefore the middle homomorphism in (\ref{eqn:homos}) is injective.

Summing up, we conclude that the middle homomorphism is an
isomorphism, and $Y_{c_1}$ is homologically equivalent to $X_{c_1}$.

Similarly, $Y_{c_2}$ is homologically equivalent to $X_{c_2}$. But
the sets $X_{c_1}$ and $X_{c_2}$ are not homologically equivalent:
the first is homologically equivalent to a sphere, and the second is
contractible. Therefore $Y_{c_1}$ and $Y_{c_2}$ are also
homologically different implying that $V_{\alpha}^{-\frac 1 {\alpha}}$ should
have a critical point in $\mathcal{U}_K$.

\subsubsection{Completing the proof of Theorem \ref{th:asymp}.a}

Since $V_{\alpha}$ has no critical points outside the convex hull of
the charges,  it is enough to consider  instead of $\RR^n$
an open  ball $B$ containing all charges.

It is easy to see  that one can cover $B$ by neighborhoods
$\mathcal{U}_k$ as in Lemma \ref{lem:choice of U} and
\ref{lem:choice of Uk}. Indeed, start from zero-dimensional Voronoi
cells $S_i$, and choose their neighborhoods $\mathcal{U}_i$
according to Lemma \ref{lem:choice of U}. Then choose compacts
$K^1_i$ in one dimensional Voronoi cells in such a way that their
union with these neighborhoods covers the intersection of the union
of all one-dimensional Voronoi cells with $B$. Choose neighborhoods
$\mathcal{U}_{K^1_i}$ of these compacts according to Lemma
\ref{lem:choice of Uk}. Then choose compact subsets $K^2_i$ of
two-dimensional Voronoi cells  in such a way that $(\cup
K^2_i)\cup(\cup\mathcal{U}_{K^1_i})$ covers the intersection of the
union of all two-dimensional Voronoi cells with $B$, and so on. At
the end the union of all selected neighborhoods will cover $B$.

Each of the neighborhoods corresponding to effective Voronoi cells
will contain one critical point of $V_{\alpha}$, and its Morse index
will be equal to the dimension of the Voronoi cell. The
neighborhoods of non-effective Voronoi cells will not contain
critical points of $V_{\alpha}$.

\subsection{Proof of Theorem \ref{th:asymp}.b}

We are looking for the critical points of the function
$\tilde{V}_{\alpha}$ defined in a linear space $N$
$$
\tilde{V}_{\alpha}=\sum\zeta_i
\tilde{\rho}_i^{-\alpha},\qquad\text{where} \quad
\tilde{\rho}_i=\dist^2(\bar x,\tilde{c}_i)+y_i^2,
$$
where $\tilde{c}_i$ are now \emph{orthogonal projections} on $N$ of
the positions $c_i$ of the charges $\zeta_j$. The claim is that the
critical points of $\tilde{V}_{\alpha}$ are in one-to-one
correspondence with the effective with respect to $N$ cells of the
Voronoi diagram of $\{\zeta_j\}$.

We can characterize the partition of $N$ by intersections with the
cells of the Voronoi diagram  only in terms of $\tilde{\rho}_i$.
Namely, a generalized Voronoi diagram in $N$ is defined as the
classical Voronoi diagram in \$\ref{ssec:defs and main conj}, with
$\rho_i$ replaced by $\tilde{\rho}_i$: a cell $S$ of a generalized
Voronoi diagram is the set of all points $\bar{x}\in N$ with the same
set $\NS(S)=\{i|\forall
k\quad\tilde{\rho}_i(\bar{x})\le\tilde{\rho}_k(\bar{x})\}$. One can
immediately see that thus defined generalized Voronoi diagram
coincides with the intersection of the original Voronoi diagram with
$N$.

It turns out that using this notation one can get
the proof of Theorem \ref{th:asymp}.b from that of Theorem \ref{th:asymp}.a by a simple
replacement of $\RR^n$ by $N$, $\rho_j$ by $\tilde{\rho}_j$, the
charges $\zeta_j$ by their projections on $N$, $\nabla{V}_{\alpha}$
by $\nabla\tilde{V}_{\alpha}$, and cells of the Voronoi diagram by
cells of the generalized Voronoi diagram. Namely, since
$\tilde{\rho}_j$ is  just the square of the distance
to $\tilde{c}_j$ (up to a constant), exactly the same formulae for
$\nabla\tilde{V}_{\alpha}$ and $\Hess\tilde{V}_{\alpha}$ hold. Since
$\tilde{\rho}_j$'s are radially symmetric, the condition that $N$
intersects the Voronoi diagram generically implies the linear
independence of $\nabla\tilde{\rho}_j$, which guarantees the second
property of $\mathcal{U}$ and $\mathcal{U}_K$, etc.

The only difference appears for the full-dimensional strata, i.e. the case
of $k=0$ in Lemma \ref{lem1:dim=k} and Lemma \ref{lem2:dim=k}: while
in Theorem \ref{th:asymp}.a these strata do not contain critical
points, in Theorem \ref{th:asymp}.b the full-dimensional strata
corresponding to strictly positive $\tilde{\rho}_j$ will have a
critical point. This point will be necessarily unique and is a local
maximum by  Lemma \ref{lem1:dim=k} and Lemma \ref{lem2:dim=k} (modified
as mentioned above).

We leave it as an exercise to check that the aforementioned
modifications of the proof of Theorem \ref{th:asymp}.a produce a
correct proof of Theorem \ref{th:asymp}.b.

\subsection{Proof of  Theorem \ref{th:origMx}}

Denote by $a^j_{\alpha}$ the number of critical points of Morse index
$j$ of $V_{\alpha}(\bar x)$. The standard potential $V_{1}(\bar x)$
is harmonic  in $\RR^3$, and, therefore, has no local maxima/minima,
i.e. $a^0_1=0$. Using the Euler characteristics one can easily check
that $a^2_{\al}-l+1=a^1_{\al}-a^0_{\al}$ for any $\al$. Therefore,
for $\al=1$ the total number of critical points
$a^2_{1}+a^1_{1}+a_{0}^1$ equals $2a_{1}^2-l+1$. By Maxwell
inequalities (\ref{Morse}) one gets $a_{1}^2\le a_{\infty}^2\le
\frac{l(l-1)}{2}$, see \S 1. Thus,
$a^2_{1}+a^1_{1}+a_{0}^1=2a_{1}^2-l+1\le l(l-1)-l+1=(l-1)^2$, exactly
Maxwell's estimate.

\section{Remarks and problems}
\label{sec:rems}

{\it Remark 1.} Main objects of consideration in this paper have a strong
resemblance with the main objects in tropical algebraic geometry.
Namely, the potential $V_{\al}(\bar x)$ resembles an actual algebraic
hypersurface while $V_{\infty}(\bar x)$ resembles its tropical limit.
Also, Voronoi diagrams are piecewise linear objects as well as
tropical curves. It is a pure coincidence?

\medskip
{\it Remark 2.} What happens in  the case of charges of different
signs? Note that in a Voronoi cell of highest dimension
corresponding to a negative charge the potential of this charge
outweighs potentials of all other charges for large $\alpha$, and
$|V_{\alpha}|^{-1/\alpha}$ converges uniformly on compact subsets of
this cell to $V_{\infty}(\bar x)$. Therefore it seems that the
function defined on the union of Voronoi cells of highest dimension
as
$$
 \tilde{V}_{\infty}(\bar x)=\operatorname{sign}\zeta_i\cdot\rho_i(x),\quad\text{if}\quad
\rho_i(x)=\min_j\rho_j(x)
$$
is responsible for the  critical points of $V_{\alpha}$ as
$\alpha\to\infty$.

\medskip
{\it Remark 3.} Theorem \ref{th:asymp}  is similar to the
results of Varchenko and Orlik-Terao on the number of critical
points for the product of powers of real linear forms and the
number of open components in the complement to the corresponding
arrangement of affine hyperplanes. Is there an appropriate result?

   \medskip
   {\it Remark 4.} Conjecturally the number of critical points of
   $V_{\al}(\bar x)$ is bounded from above by the number of effective
   Voronoi cells in the corresponding Voronoi diagram. The number of all
   Voronoi cells in Voronoi diagrams in $\RR^n$ with $l$ sites has a nice
   upper bound. What is the upper bound for the number of effective
   Voronoi cells? Is it the same as for all Voronoi cells?
     %IN GENERAL, LET US CALL A CONFIGURATION ALL
      %      Voronoi cells OF VORONOI DIAGRAM OF WHICH ARE EFFECTIVE an
       %     $M$-configuration. Problem: DeSCRIBE $M$-CONFIGURATIONS OR AT
        %    LEAST FIND SERIES OF EXAMPLES. A WILD GUESS THAT THIS MIGHT BE
         %   RELATED TO THOMPSON'S CLASSICAL PROBLEM ON EQUILIBRIUM
          %  DISTRIBUTION OF UNIT CHARGES ON $S^2$ WHICH ATIYAH ADVERTISED
           % RECENTLY. MAYBE ANY GENERIC CONFIGURATION CLOSE TO AN
            %EQUILIBRIUM IS AN $M$-CONF???

\medskip
{\it Remark 5.} Many  statements in the paper are valid if one
substitutes the potential $r^{-\al}$ of a unit charge located at the origin
by more or less any concave function $\psi(r)$ of the radius in $\RR^n$.
To what extent the above results and conjecture can be generalized for
$\psi(r)$-potentials?

\medskip
{\it Remark 6.} The initial hope in settling
Conjecture~\ref{conj:main} was related to the fact that in our
numerical experiments for a fixed configuration of charges the
number of critical points of $V_{\al}(\bar x)$ was a  {\bf
nondecreasing} function of $\alpha$. Unfortunately this monotonicity
turned out to be wrong in the most general formulation: the number
of critical points of a restriction of a potential to a line is not
a monotonic function of $\alpha$.
\begin{example}

The potential\; $
V_{\al}(x)=[(x+30)^2+25]^{-\alpha}+[(x+20)^2+49]^{-\alpha}+$ $
[(x+2)^2+144]^{-\alpha} +[(x-20)^2+49]^{-\alpha} +
[(x-30)^2+25]^{-\alpha} $ has three critical points for
$\alpha=0.1$, seven critical points for $\alpha=0.2$, again three
critical points for $\alpha=0.3$, and again seven critical points as
$\alpha=1.64$, and nine critical points for $\alpha\ge 1.7$.
\end{example}

Existence of  such an example for the potential itself (and not of
its restriction) is unknown.

\section{Appendix: James C. Maxwell on  points of equilibrium}
\label{sec:append}

In his monumental Treatise \cite{Maxwell} Maxwell has foreseen the
development of several mathematical disciplines. In the passage
which we have the pleasure to present to the  readers his
arguments are that of Morse theory developed at least 50 years
later. He uses the  notions of periphractic number, or, degree of
periphraxy which is the rank of $H_{2}$ of a domain in $\RR^3$
defined as the number of interior surfaces bounding the domain and
the notion of cyclomatic number, or, degree of cyclosis which is
the rank $H_{1}$ of a domain in $\RR^3$ defined as the number of
cycles in a curve obtained by a homotopy retraction of the domain
(none of these notions rigorously existed then). (For definitions of these notions see \cite{Maxwell}, section 18.) Then he actually
proves Theorem \ref{MK} of \S \ref{sec:intro} usually attributed
to M.~Morse. Finally in Section [113] he makes the following claim.
\medskip

" To determine the number of the points and lines of equilibrium,
let us consider the surface or surfaces for which the potential is
equal to $C$, a given quantity. Let us call the regions in which
the potential is less than $C$ the negative regions, and those in
which it is greater than $C$ the positive regions. Let $V_{0}$ be
the lowest and $V_{1}$ the highest potential existing in the
electric field. If we make $C=V_{0}$ the negative region will
include only one point or conductor of the lowest potential, and
this is necessarily charged negatively. The positive region
consists of the rest of the space, and since it surrounds the
negative region it is periphractic.

If we now increase the value of $C$, the negative region will
expand, and new negative regions will be formed round negatively
charged bodies. For every negative region thus formed the
surrounding positive region acquires one degree of periphraxy.

As the different negative regions expand, two or more of them may
meet at a point or a line. If $n+1$ negative regions meet, the
positive region loses $n$ degrees of periphraxy, and the point or
the line in which they meet is a point or line of equilibrium of
the $n$th degree.

When $C$ becomes equal to $V_{1}$ the positive region is reduced
to the point or the conductor of highest potential, and has
therefore lost all its periphraxy. Hence, if each point or line of
equilibrium counts for one, two, or $n$, according to its degree,
the number so made up by the points or lines now considered will
be less by one than the number of negatively charged bodies.

There are other points or lines of equilibrium which occur where
the positive regions become separated from each other, and the
negative region acquires periphraxy. The number of these, reckoned
according to their degrees, is less by one than the number of
positively charged bodies.

If we call a point or line of equilibrium positive when it is the
meeting place of two or more positive regions, and negative when
the regions which unite there are negative, then, if there are $p$
bodies positively and $n$ bodies negatively charged, the sum of
the degrees of the positive points and lines of equilibrium will
be $p-1$, and that of the negative ones $n-1$. The surface which
surrounds the electrical system at an infinite distance from it is
to be reckoned as a body whose charge is equal and opposite to the
sum of the charges of the system.

But, besides this definite number of points and lines of
equilibrium arising from the junction of different regions, there
may be others, of which we can only affirm that their number must
be even. For if, as any one of the negative regions expands, it
becomes a cyclic region, and it may acquire, by repeatedly meeting
itself,  any number of degrees of cyclosis, each of which
corresponds to the point or line of equilibrium at which the
cyclosis was established. As the negative region continues to
expand till it fills all space, it loses every degree of cyclosis
it has acquired, a becomes at last acyclic. Hence there is a set
of points or lines of equilibrium at which cyclosis is lost, and
these are equal in number of degrees to those at which it is
acquired.

If the form of the charged bodies or conductors is arbitrary, we
can only assert that the number of these additional points or
lines is even, but if they are charged points or spherical
conductors, the number arising in this way cannot exceed
$(n-1)(n-2)$ where $n$ is the number of bodies*.

*\{I have not been able to find any place where this result is
proved.\}.

\medskip
We finish the paper by mentioning that the last remark was added by
J.~J.~Thomson in 1891 while proofreading the third (and the
last) edition of Maxwell's book. Adding the above numbers of
obligatory  and additional critical points one arrives at the
conjecture \ref{Mx} which was the starting point of our paper.

\end{document}